\documentclass[fleqn,10pt]{wlscirep}
\usepackage[utf8]{inputenc}
\usepackage[T1]{fontenc}
\usepackage{overpic}
\usepackage{graphicx}
\usepackage{tikz}
\usepackage{caption}
\usepackage{xcolor}
\usepackage{multirow}
\usepackage{booktabs}
\usepackage{caption}
\usepackage{float}
\usepackage{textcomp}

\definecolor{mygreen}{HTML}{008000}
\definecolor{myorange}{HTML}{FFA500}

\title{The Impact of Meteorological Factors on Crop Price Volatility in India: Case studies of Soybean and Brinjal}

\author[1]{Ashok Kumar}
\author[2]{Abbinav Sankar Kailasam}
\author[3, 4]{Anish Rai}
\author[5]{Manya Khanna}
\author[6]{Sudeep Shukla}
\author[3]{Sourish Das}
\author[1,*]{Anirban Chakraborti}

\affil[1]{School of Computational \& Integrative Sciences, Jawaharlal Nehru University, New Delhi-110067, India}
\affil[2]{School of Computing and Data Science, Sai University, Chennai-603104, Tamil Nadu, India}
\affil[3]{Chennai Mathematical Institute, Chennai-603103, Tamil Nadu, India}
\affil[4]{Department of Physics, National Institute of Technology Sikkim-737139, Sikkim, India}
\affil[5]{Center for Creative Leadership, Gurugram, India}
\affil[6]{AI 4 Water LTD, Orpington, BR6 9QX United Kingdom}

\affil[*]{anirban@jnu.ac.in}

\begin{abstract}
Climate is an evolving complex system with dynamic interactions and non-linear feedback mechanisms, shaping environmental and socio-economic outcomes. Crop production is highly sensitive to climatic fluctuations (and many other environmental, social and governance factors). This paper studies the price volatility of agricultural crops as influenced by meteorological variables, which is critical for agricultural planning, sustainable finance and policy-making. As case studies, we choose the two Indian states: Madhya Pradesh (for Soybean) and Odisha (for Brinjal/Eggplant). We employ an Exponential Generalized Autoregressive Conditional Heteroskedasticity (EGARCH) model to estimate the conditional volatility of the log returns from 2012 to 2024. We further explore the cross-correlations between price volatility and the meteorological variables followed by a Granger-causal test to analyze the causal effect of meteorological variables on the volatility. The Seasonal Auto-Regressive Integrated Moving Average with Exogenous Regressors (SARIMAX) and Long Short-Term Memory (LSTM) models are implemented as simple machine learning models of price volatility with meteorological factors as exogenous variables. Finally, to capture spatial dependencies in volatility across districts, we extend the analysis using a Conditional Autoregressive (CAR) model to construct monthly volatility surfaces that reflect both local price risk as well as geographic dependence. We believe, this paper will illustrate the usefulness of simple machine learning models in agricultural finance, and help the farmers to make informed decisions by considering climate patterns and making beneficial decisions with regard to crop rotation or allocations. In general, incorporating meteorological factors to assess agricultural performance could help to understand and reduce price volatility and possibly lead to economic stability.
\end{abstract}

\begin{document}

\flushbottom
\maketitle
\thispagestyle{empty}

\section{Introduction}

Systems--whether natural, like climate and hydrology, or artificial, like financial markets--comprise of numerous independent components whose intricate interactions give rise to emergent behaviours~\cite{chakraborti2020emerging,chakraborti2020phase,rai2024identifying}. These behaviours often cannot be predicted by examining individual parts in isolation~\cite{chakrabarti2023data,kwapien2012physical}. Such systems are prone to extreme events with far-reaching consequences across natural, economic, and social environment. Amongst these, climate stands out as an evolving complex system characterized by dynamic interactions and non-linear feedback mechanisms that shape environmental and socio-economic outcomes~\cite{rind1999complexity}. Climate change, being one of the most pressing challenges today, exemplifies the multifaceted nature of such systems. India, in particular, ranks sixth among countries most affected by its impact~\cite{adil2025climate}. Given the complexity and feedback-driven interactions characteristic of climate systems, employing advanced analytical approaches is essential for a comprehensive understanding. For instance, dynamical systems theory has been applied to study climate variability in India, revealing significant dynamic shifts during strong El Niño and La Niña events and uncovering larger patterns within climate data\cite{john2024recurrence}. Such methodologies provide deeper insights into the intricate behaviors of climate systems and their broader implications.

Agriculture--especially in developing countries like India--is extremely sensitive to climate fluctuations. Even minor shifts in temperature or precipitation can disrupt planting cycles, reduce yields, and destabilize supply chains, resulting in sharp market fluctuations and increased price volatility~\cite{kumar2014climate,Zhao2017,Burney2014,Burney2024}. For farmers whose primary source of income is crop yield and sales, agricultural price fluctuations pose severe risks to their livelihood, and threaten food security~\cite{Agricultural,Carleton2017,Burney2014}. In India, where agriculture is both, an important economic sector and a vital source of employment, understanding and predicting the dynamics of crop price volatility is essential. Although traditional research has shown how average weather conditions affect crop yields or overall prices~\cite{Agricultural}, volatility of crop prices due to climate variability has not been explored much. 

This study aims to address this gap by applying a complex systems approach to analyze agricultural price volatility. Crop prices are treated as a financial time series to analyze their volatility characteristics, with log returns and squared log returns computed to estimate conditional volatility. Subsequently, the effect of meteorological variables (temperature and precipitation) on crop price volatility is evaluated, using a methodology that allows for the examination of non-linear and asymmetric effects of climate change on agricultural markets. Two distinct crops and regions have been analyzed: soybean in Madhya Pradesh, an export-oriented \textit{kharif} crop, and brinjal(eggplant) in Odisha, a domestically consumed vegetable grown all year. Soybean in Madhya Pradesh was chosen due to Madhya Pradesh's dominant role in India's soybean sector--contributing approximately 42\% of national production as of 2023-24\cite{timesofindia2024soybean}. The state, often referred to as the "Soybean State"\cite{timesofindia2024soybean}, has the largest area under soybean cultivation, approximately 5.35 million hectares\cite{srivastava2014analysis}. Brinjal(eggplant) cultivation in Odisha was chosen due to the state's status as the second-largest producer of the vegetable in India\cite{isaaa_pk35_2008}. While West Bengal leads in overall production, its highly diverse climate poses challenges for consistent agricultural studies\cite{das2023climate}. In contrast, Odisha offers relatively uniform meteorological conditions across the state, making it a more suitable and controlled environment for focused analysis. A combination of econometric (EGARCH) and machine learning models (SARIMAX and LSTM) is employed to investigate whether and how precipitation and maximum temperature affect crop price volatility, and whether these effects can be accurately forecasted. Further, to enrich the spatial understanding of agricultural price volatility, a Conditional Autoregressive (CAR) model has been applied to construct monthly volatility surfaces across Madhya Pradesh and Odisha. This spatially explicit modelling provides a granular view of volatility distribution, offering a valuable tool for assessing local-level price risk. By integrating high-resolution meteorological data with market prices of the two crops, the study seeks to anticipate risk and offer insights into how tailored financial models can mitigate the socio-economic challenges posed by climate-driven agricultural volatility. The findings of the study highlight the extent to which meteorological variables influence agricultural commodity prices. This analysis aims to elucidate how climate variability translates into price volatility offering insights that are both theoretically and practically relevant for policy-making and risk management.

\section{Literature Review}

The fluctuations in crop supply directly influence market prices, leading to increased price volatility that can destabilize local economies and threaten the financial well-being of smallholder farmers\cite{Modelling,Burney2014,Burney2024}. Extreme weather events, like precipitation anomalies, have a pronounced impact on economic inequality in nations that rely heavily on agriculture. Research shows that in these nations a 1.5 standard deviation increase in precipitation relative to average levels affects the low-income groups 35 times more in countries where agricultural employment is high (for example, Africa), compared to those with minimal agricultural reliance. This shows that anomalies in rainfall and the level of agricultural dependence are important factors to consider while evaluating the negative impacts of climate change on vulnerable groups\cite{doi:10.1073/pnas.2203595119}. Climate change also disrupts broader environmental systems. For instance, changes in temperature can offset the expected benefits of increased rainfall on nitrogen runoff. Additionally, it also disrupts the movement of nitrogen from land to aquatic systems. The intricate relationship between environmental sustainability, economic inequality, and climate change is made clear by these revelations\cite{doi:10.1073/pnas.2220616120}.

Looking ahead, climate projections suggest that crop yields could be reduced to a large extent if global temperatures rise beyond ideal thresholds for crops--29°C for corn, 30°C for soybean, and 32°C for cotton. The U.S., which produces 41\% of global maize and 38\% of soybean, may experience production declines of up to 82\%\cite{doi:10.1073/pnas.0906865106}. The development of heat-tolerant crops is therefore vital since warming of 2°C and 4°C would likely increase the variability of maize yields, threatening food security and the stability of the grain trade, particularly for the 800 million people living in extreme poverty\cite{doi:10.1073/pnas.1718031115}. In West Bengal for instance, transitions toward non-food grains are being influenced by temperature and humidity, pointing to the complex and localized impact of climate change\cite{PARIA2022100499}. Thus, India's agricultural sector must adapt its cropping patterns for sustainable growth.

Although agricultural price volatility has historically been studied through market-centric frameworks, there is a growing recognition of the influence of meteorological drivers. Before exploring advancements in analytic tools and methodologies, it is important to first revisit the classical economic theories that have shaped our understanding of agricultural price volatility. In the early eighties, under the influence of the "Washington Consensus", trade liberalization, reduced state intervention, and reliance on international markets for food security was advocated for\cite{Gerard2012Agricultural}. The theoretical foundation of such policies were rooted in welfare economics, which states that free markets guarantee maximum welfare outcomes given economic agents are rational and perfect information exists\cite{Fama1965a,Samuelson1965}. Traditional economic models often attribute agricultural price volatility to exogenous shocks affecting supply, such as climate change or pests, coupled with inelasticity of demand. These shocks are typically represented as Gaussian disturbances in applied models\cite{Gerard2012Agricultural}. The recommended solution to overcome price instability is trade and private storage. This is because operators are incentivised to buy when prices are low and sell when prices are high, leading to market enlargement and stability through the "law of large numbers"\cite{TyersAnderson1992, Vanzetti1998}. However, these approaches failed to predict the global food price crisis of 2006-2008 and 2010-2011, raising questions about the understanding of commodity price determinants and the validity of the "rational expectations hypothesis"\cite{Gerard2012Agricultural}. Literature demonstrates that markets do not always converge to an equilibrium. This highlights the role of endogenous factors and market imperfections while understanding agricultural price volatility. Newer models such as Generalized Autoregressive Conditional Heteroskedasticity (GARCH)\cite{Bollerslev1986} and Exponential Generalized Autoregressive Conditional Heteroskedasticity (EGARCH)\cite{Nelson1991} have been used to analyze time series data. These models capture asymmetric responses to shocks, making them particularly suitable for agricultural contexts, where weather-induced price shifts can be sharp and skewed. Research employing GARCH-MIDAS models have revealed that temperature anomalies and irregular precipitation are crucial determinants of food price volatility\cite{Modelling,Burney2014,Burney2024}. These insights underscore the need to integrate meteorological variables into volatility modelling. Climate-induced volatility, when appropriately modeled, enables more effective agricultural risk management strategies.   

Predictive models that incorporate meteorological variables offer a strategic response to risks posed by climate vulnerability. By embedding temperature and precipitation data, forecasting tools can assist policymakers and financial institutions in designing resilient crop insurance schemes and other mitigation mechanisms\cite{Mahawan2022HybridAA,Zhao2017,Carleton2017}. Among various models, hybrid approaches such as ARIMAX-LSTM models outperform traditional methods, achieving lower mean absolute percentage error (MAPE) values thereby enhancing the accuracy of price forecasts and risk mitigation strategies\cite{Modelling,Burney2024}. They capture both linear and non-linear dependencies between climate variables and price volatility. A complex systems approach, as utilised in this study, addresses limitations of traditional linear models by framing agricultural markets as systems with interacting parts--farmers, markets, weather patterns, and policy mechanisms--whose collective behaviour exhibits emergent properties.   

This study contributes to the literature by applying these hybrid methodologies to two key agricultural commodities in India: soybean in Madhya Pradesh and brinjal(eggplant) in Odisha. These represent contrasting market contexts-- soybean as an export-oriented kharif oilseed, and brinjal(eggplant) as a domestically consumed, non-seasonal vegetable~\cite{prashnani2024towards,agarwal2013soybean,ghosh2022eggplant,rao2010climate}. This allows for a comparative analysis of globally traded versus locally consumed commodities. The EGARCH model is used to estimate volatility, SARIMAX and LSTM models are employed to forecast price dynamics by incorporating meteorological variables like maximum temperature and precipitation. Tersely, this paper contributes to the evolution of agricultural economics by shifting the analytical lens from static, exogenous models, to dynamic climate-sensitive frameworks grounded in complex adaptive systems theory. 

\section{Data}

\subsection{Description}
The Government of India's Directorate of Marketing and Inspection operates the AGMARKNET web portal, which plays a crucial role in disseminating agricultural market information, including arrivals and prices of various agricultural commodities across India. The portal collects data from local agricultural markets through a specially designed application called "Agmark," enabling farmers, traders, and researchers to access real-time market trends and insights. The website for accessing the corresponding data is: AGMARKNET\cite{agmarknet}. By providing reliable and transparent information, AGMARKNET aims to support informed decision-making among stakeholders in the agricultural sector, ultimately fostering improved market practices and facilitating effective price discovery. We have taken monthly data from eighteen districts in Madhya Pradesh and six districts in Odisha, covering the period from Jan-2012 to Oct-2024. The districts were considered based on the data availability for this period.

The monthly data for precipitation and maximum temperature, measured at 2 meters above ground level from Jan-2012, to Oct-2024, corresponding to the crop price data for brinjal(eggplant) in Odisha and soybean in Madhya Pradesh, was sourced from NASA's POWER Project (Prediction Of Worldwide Energy Resources) and the website for the corresponding the data is NASAPOWER\cite{nasapower}. This data service provides global meteorological and solar information to support applications in renewable energy, sustainable agriculture, and climate-related projects. Powered by NASA's Earth science satellite observations and reanalysis products, including MERRA-2 (Modern-Era Retrospective analysis for Research and Applications) and CERES (Clouds and the Earth’s Radiant Energy System), the POWER project offers a comprehensive range of data for various sectors. The project primarily focuses on delivering solar irradiance data to optimize solar energy systems, alongside climate variables like temperature, precipitation, humidity, and wind speed, which are critical for crop modelling and weather monitoring in agriculture. It provides global data at high spatial resolution, with options for daily to annual time steps. This data can be accessed via an easy-to-use web interface or programmatically through an API, enabling customized dataset retrieval for specific locations and time frames.

\subsection{Preprocessing, Exploratory Analysis and Visualization}

\begin{figure}[H]
    \centering
    \begin{tikzpicture}
        \node (image) at (0,0) {\includegraphics[width=0.75\linewidth]{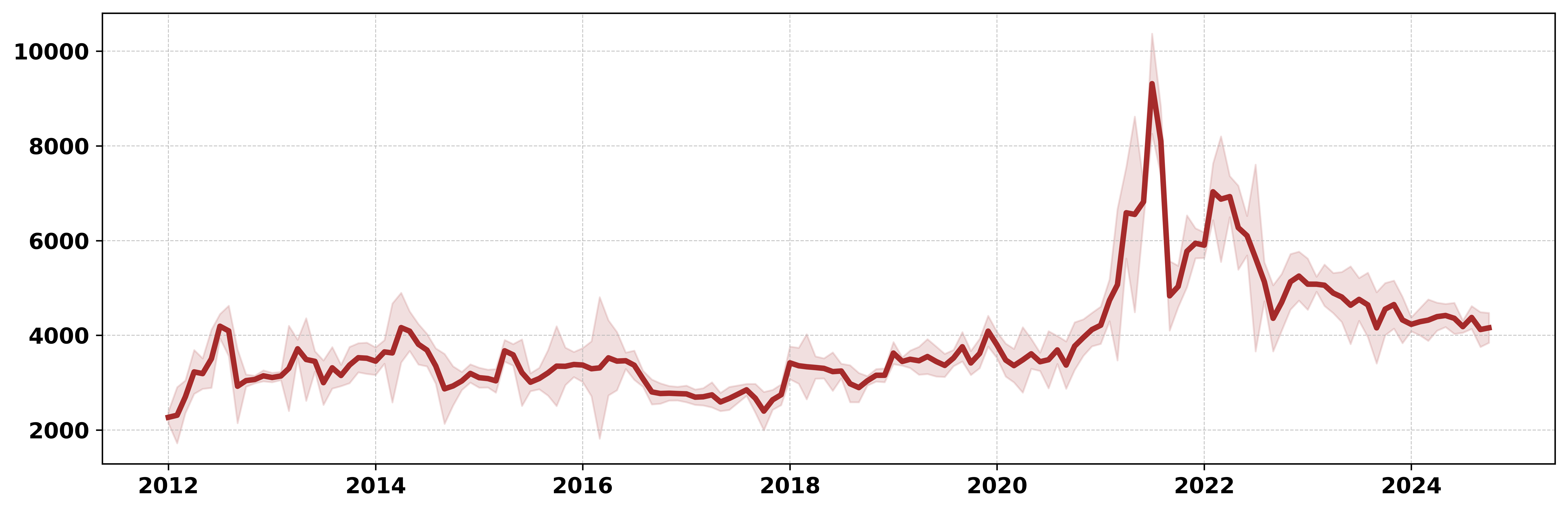}};
        \node[rotate=90] at (-7,0) {Price (INR/Quintal)};
    \end{tikzpicture}
    \begin{tikzpicture}
        \node (image) at (0,0) {\includegraphics[width=0.75\linewidth]{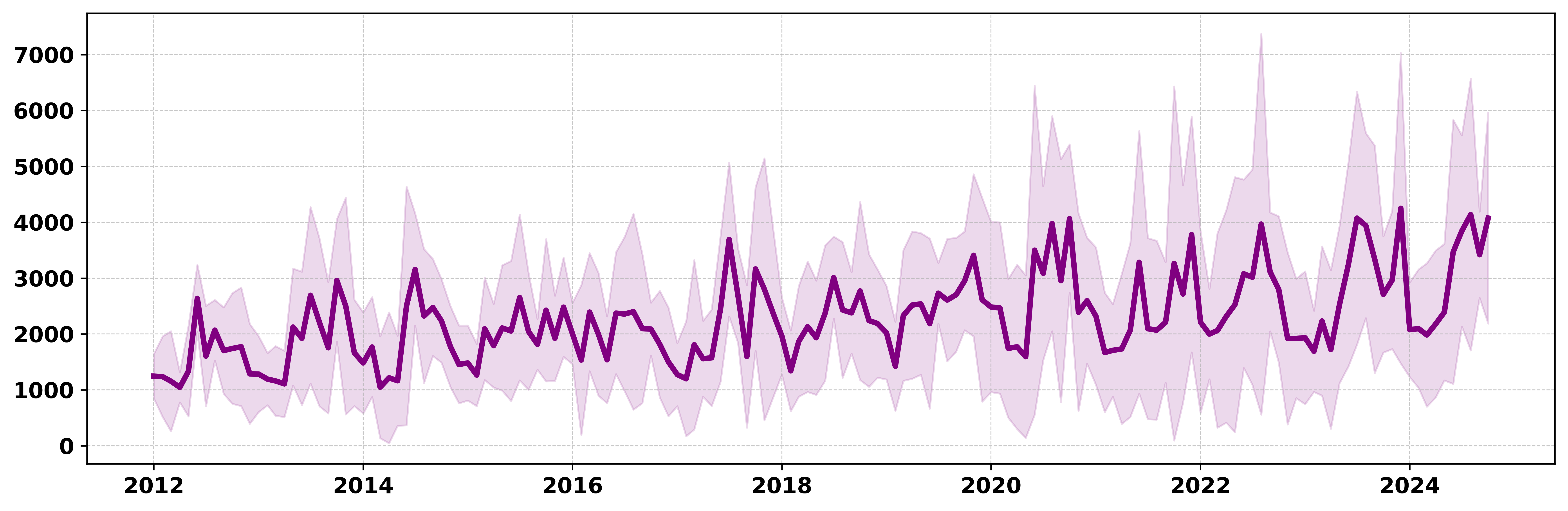}};
        \node[rotate=90] at (-7,0) {Price (INR/Quintal)};
        \node[rotate=0] at (0,-2.5) {Year};
    \end{tikzpicture}

    \captionsetup{justification=justified} 
    \caption{\textbf{Monthly Closing Price.} The brown and purple lines represent the monthly closing price of soybean and brinjal from 2012 to 2024 respectively while the shaded regions indicate the confidence intervals, calculated as $\pm 2$ standard deviation across districts, capturing the variability in prices within each state. (1 Quintal $\ensuremath{\equiv}$ 100 kg).}
    \label{fig:crop_prices}
\end{figure}

To perform a state-level analysis, the monthly closing prices for each district were averaged to derive a single representative price time series for soybean in Madhya Pradesh and brinjal in Odisha. The resulting monthly prices are shown in Fig.~\ref{fig:crop_prices}. 

The monthly closing prices for soybean show a relatively stable trend from 2012-2019, with prices largely ranging between 3000 and 4000 INR/Quintal. A significant price surge can be observed from 2020 onward peaking sharply in 2021 and early 2022, rising almost above 8000 INR/quintal. This most likely reflects the global market disruptions caused by the COVID-19 pandemic. Post 2022, prices gradually decline and show a stable pattern around 4000 INR/Quintal. Brinjal prices on the other hand, display a more volatile and erratic pattern. Prices generally range between 1000-3000 INR/Quintal from 2012-2019. Post this, an upward shift can be seen, where prices range between 4000-5000 INR/Quintal. The plot reflects the sensitivity of brinjal to localized factors like perishability, weather events, or limited post-harvest infrastructure, making its prices regionally dependent.

\begin{figure}[H]
    \centering
    \begin{tikzpicture}
        \node (image) at (0,0) {\includegraphics[width=0.75\linewidth]{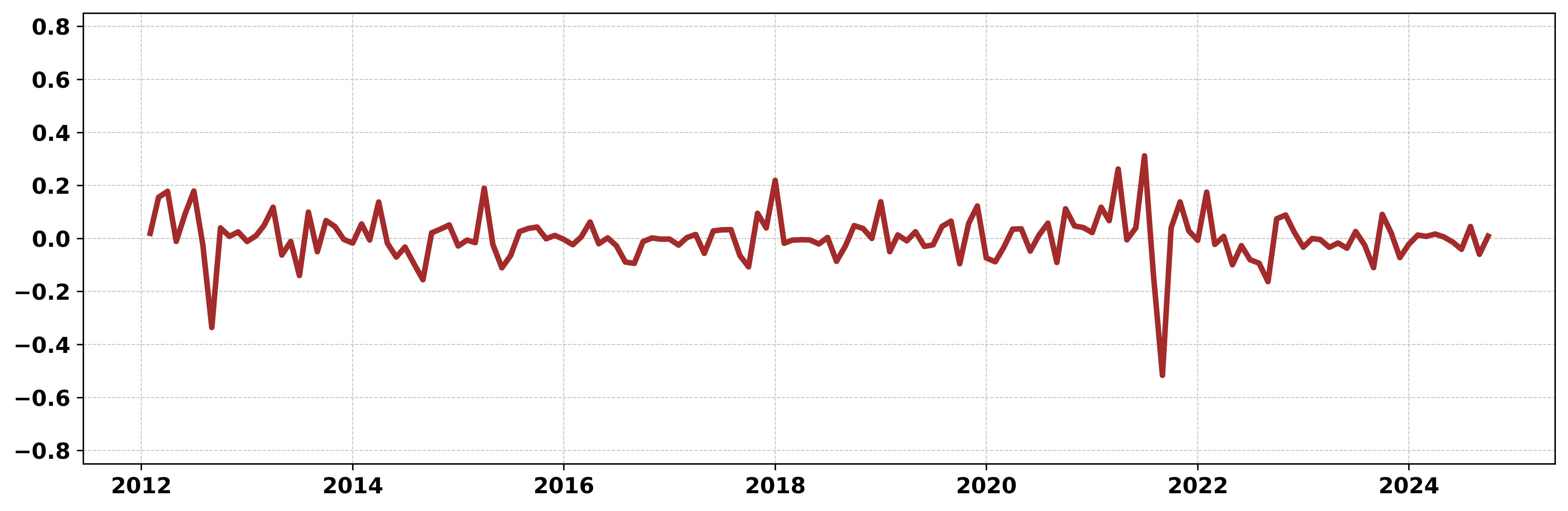}};
        \node[rotate=90] at (-7,0) {Log Returns};
        \end{tikzpicture}
    \begin{tikzpicture}
        \node (image) at (0,0) {\includegraphics[width=0.75\linewidth]{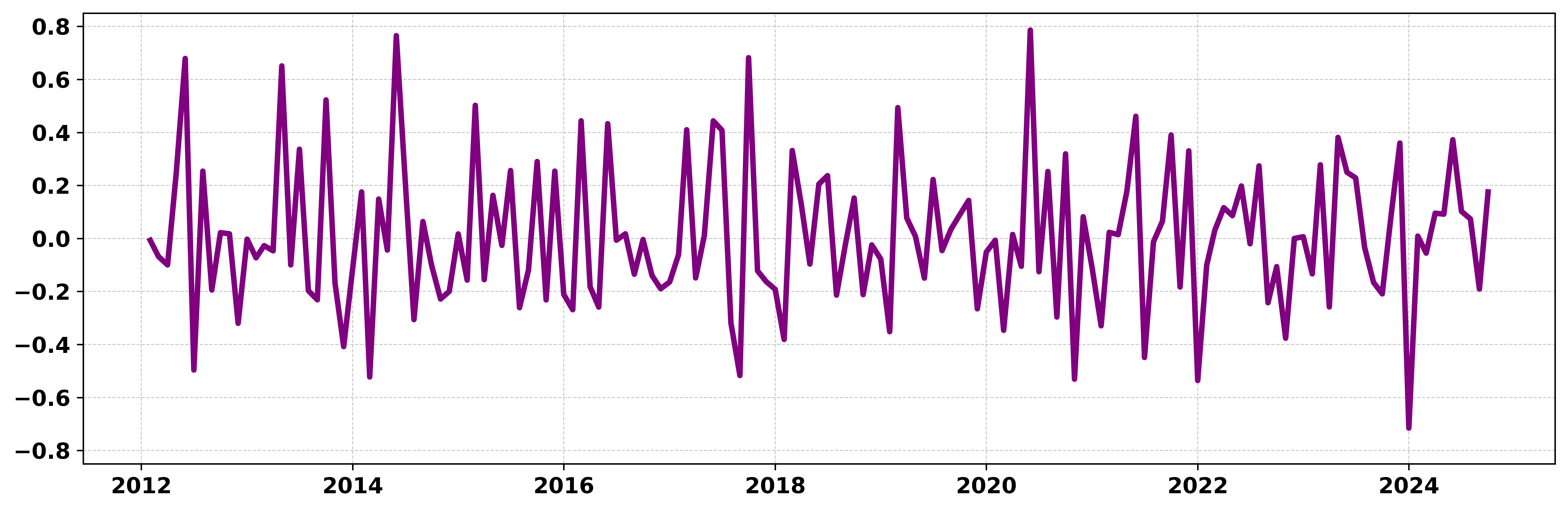}};
        \node[rotate=90] at (-7,0) {Log Returns};
        \node[rotate=0] at (0,-2.5) {Year};
    \end{tikzpicture}

    \captionsetup{justification=justified} 
    \caption{\textbf{Monthly Log Returns}. The blue and purple lines represent the monthly log returns of soybean and brinjal prices respectively.}
    \label{fig:log_returns} 
\end{figure}

Following this, the log returns of the crop prices is computed as:
\[
r_t = \ln\Bigg(\frac{P_t}{P_{t-1}}\Bigg) = \ln (P_t) - \ln (P_{t-1}),
\]
where \( r_t \) denotes the log returns at time \( t \), while \( P_t \) and \( P_{t-1} \) represent the agricultural commodity prices at times \( t \) and \( t-1 \), respectively. Figure~\ref{fig:log_returns} represents the log returns of monthly prices for soybean and brinjal respectively, measuring relative price changes over time while normalising variance. It is particularly useful for estimating volatility and measuring risk in commodity markets.

Log returns highlight potential market shocks over time. In the above figure, log returns for soybean show relatively low and stable price fluctuations with a few spikes--around 2012-2013 and early 2022--indicating brief but sharp changes in prices likely driven by external shocks. In contrast, log returns of brinjal prices show frequent and larger fluctuations, reflecting higher price volatility. These patterns suggest that brinjal prices are more unstable and reactive to short-term disturbances compared to soybean, underlying the greater market risk associated with perishable commodities. 

\begin{figure}[H]
    \centering
    \begin{tikzpicture}
        \node (image) at (0,0) {\includegraphics[width=0.75\linewidth]{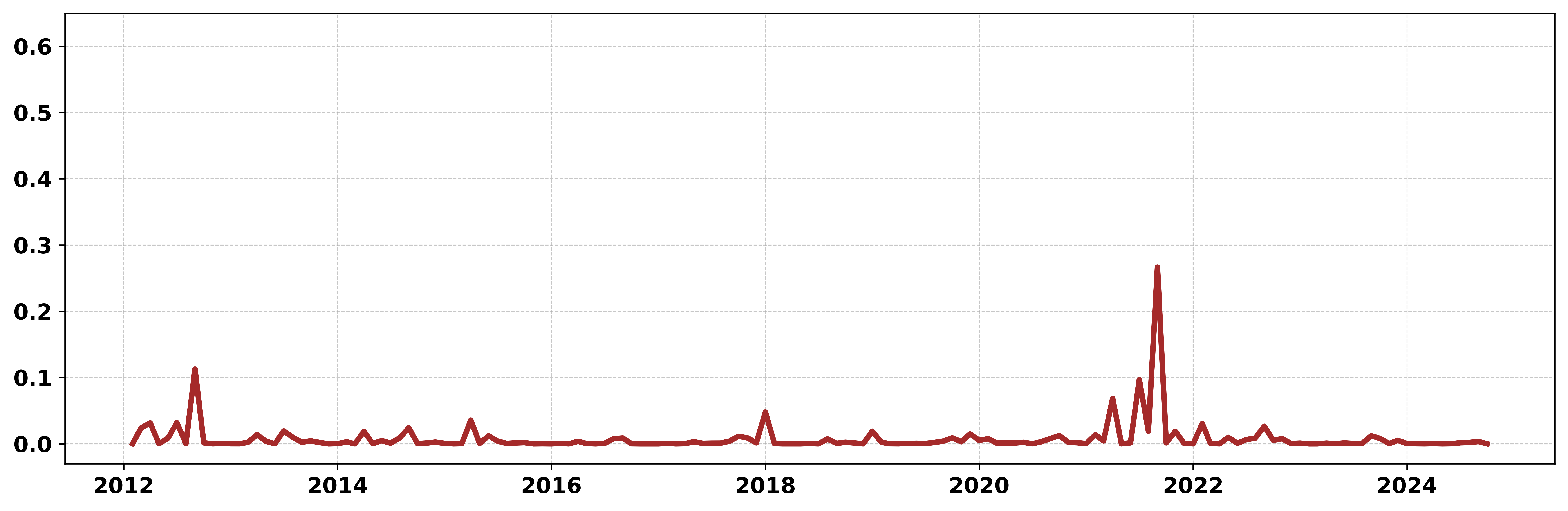}};
        \node[rotate=90] at (-7,0) {Squared Log Returns};
        \end{tikzpicture}
    \begin{tikzpicture}
        \node (image) at (0,0) {\includegraphics[width=0.75\linewidth]{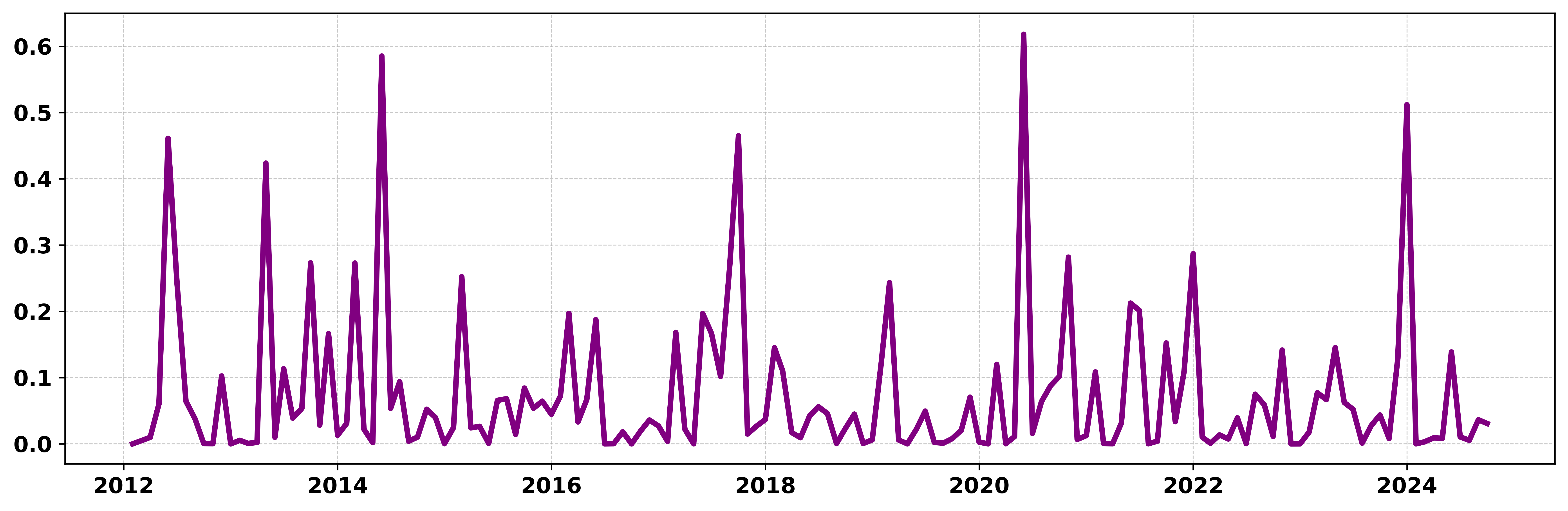}};
        \node[rotate=90] at (-7,0) {Squared Log Returns};
        \node[rotate=0] at (0,-2.5) {Year};
    \end{tikzpicture}

    \captionsetup{justification=justified} 
    \caption{\textbf{Monthly Squared Log Returns}. The brown and purple lines represent the monthly squared log returns of soybean and brinjal prices respectively.}
    \label{fig:squared_log_returns} 
\end{figure}

Figure~\ref{fig:squared_log_returns} represents the squared log returns of the monthly prices for soybean and brinjal respectively. The squared log returns exaggerate large spikes in price changes, making periods of high volatility more visually prominent. The plots show that the price volatility of brinjal is larger than soybean due to idiosyncratic factors such as localized supply-demand fluctuations and market structure.

\begin{figure}[H]
\centering
\begin{tikzpicture}
  \node (img1) at (0,0) {
    \includegraphics[width=0.4\linewidth]{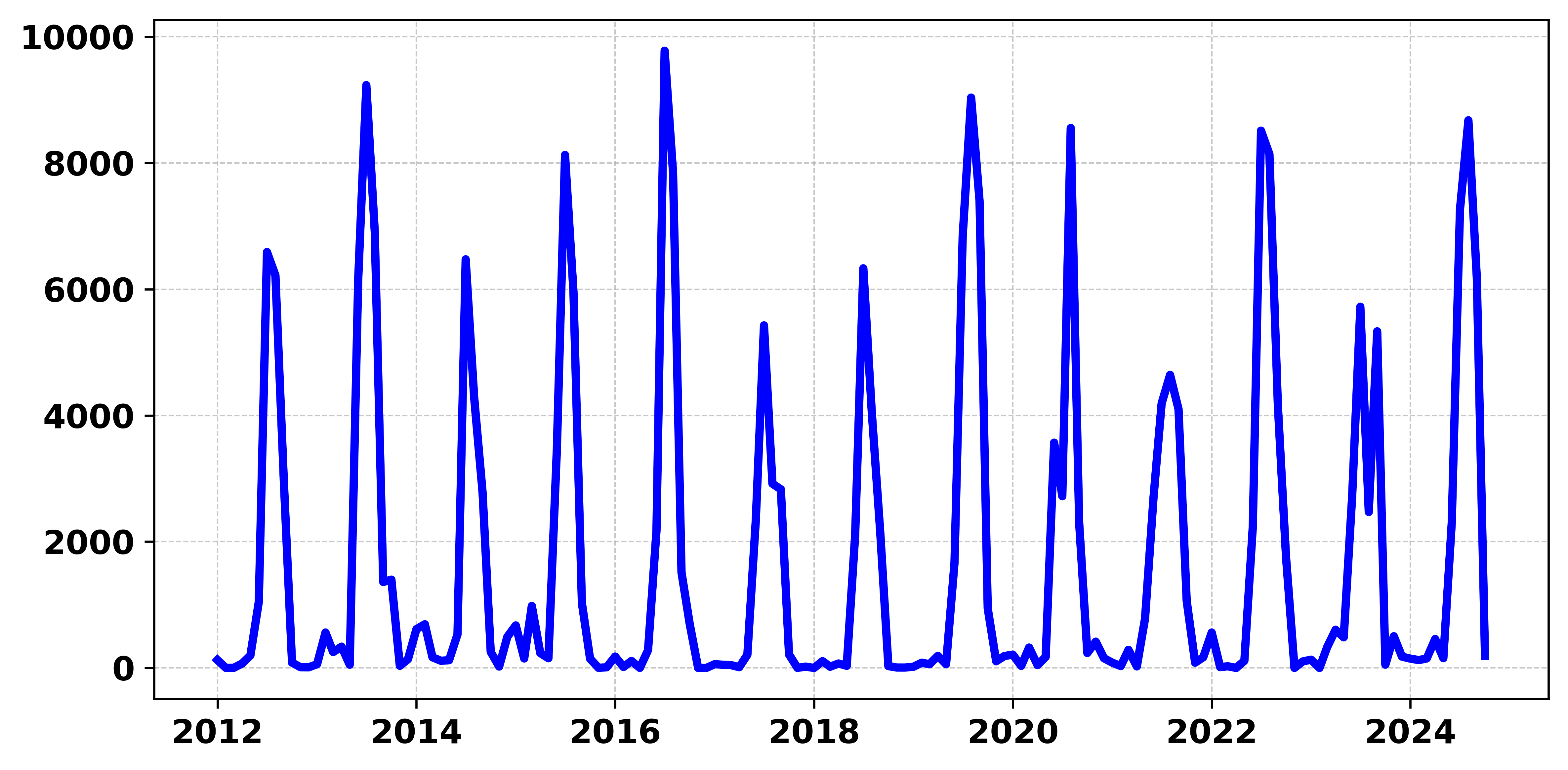}};
  \node[rotate=90] at (-4.5,0) {Precipitation (mm)};
  \node (img2) at (8,0) {
    \includegraphics[width=0.4\linewidth]{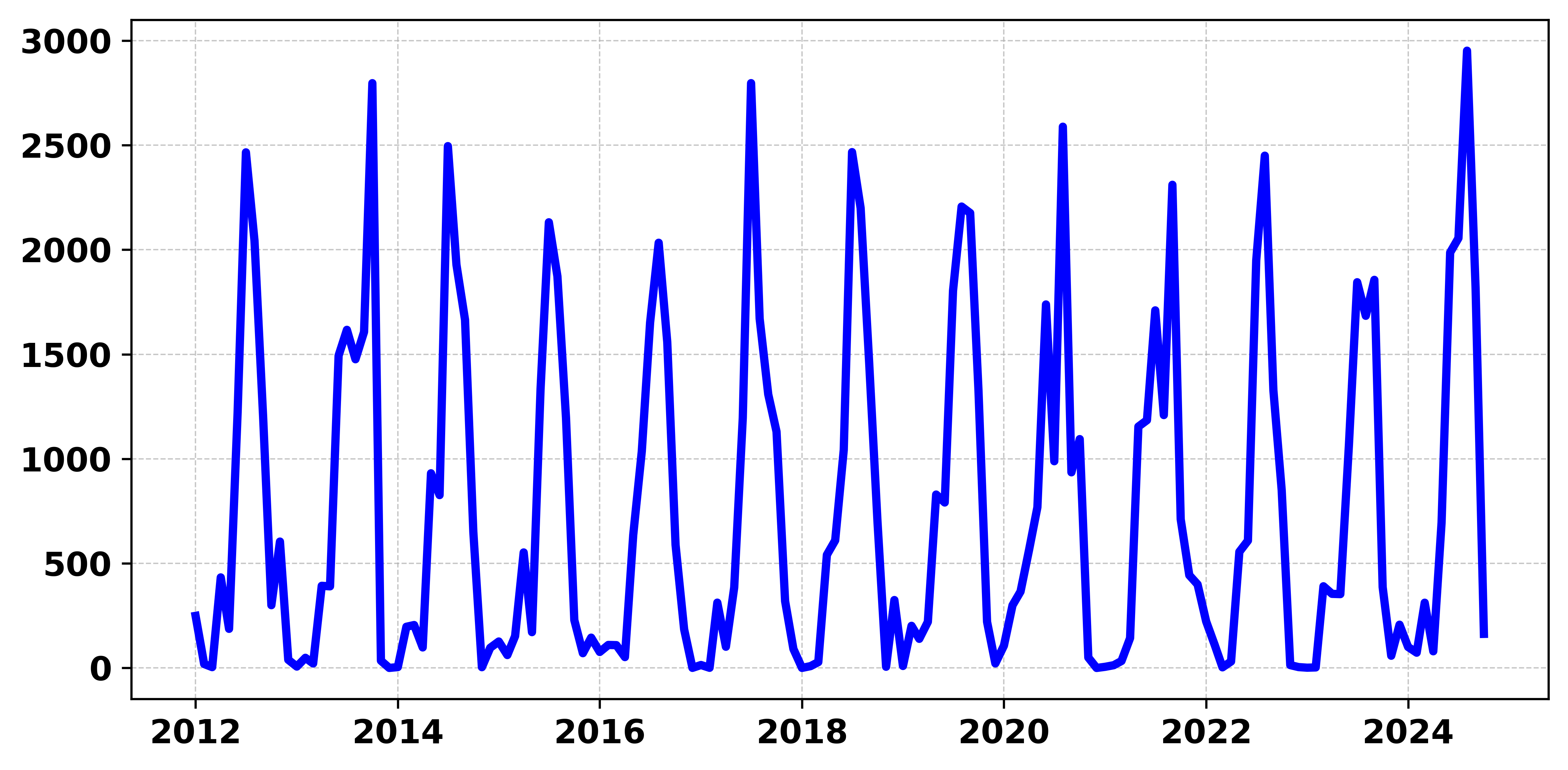}};
\end{tikzpicture}

\begin{tikzpicture}
  \node (img3) at (0,0) {
    \includegraphics[width=0.4\linewidth]{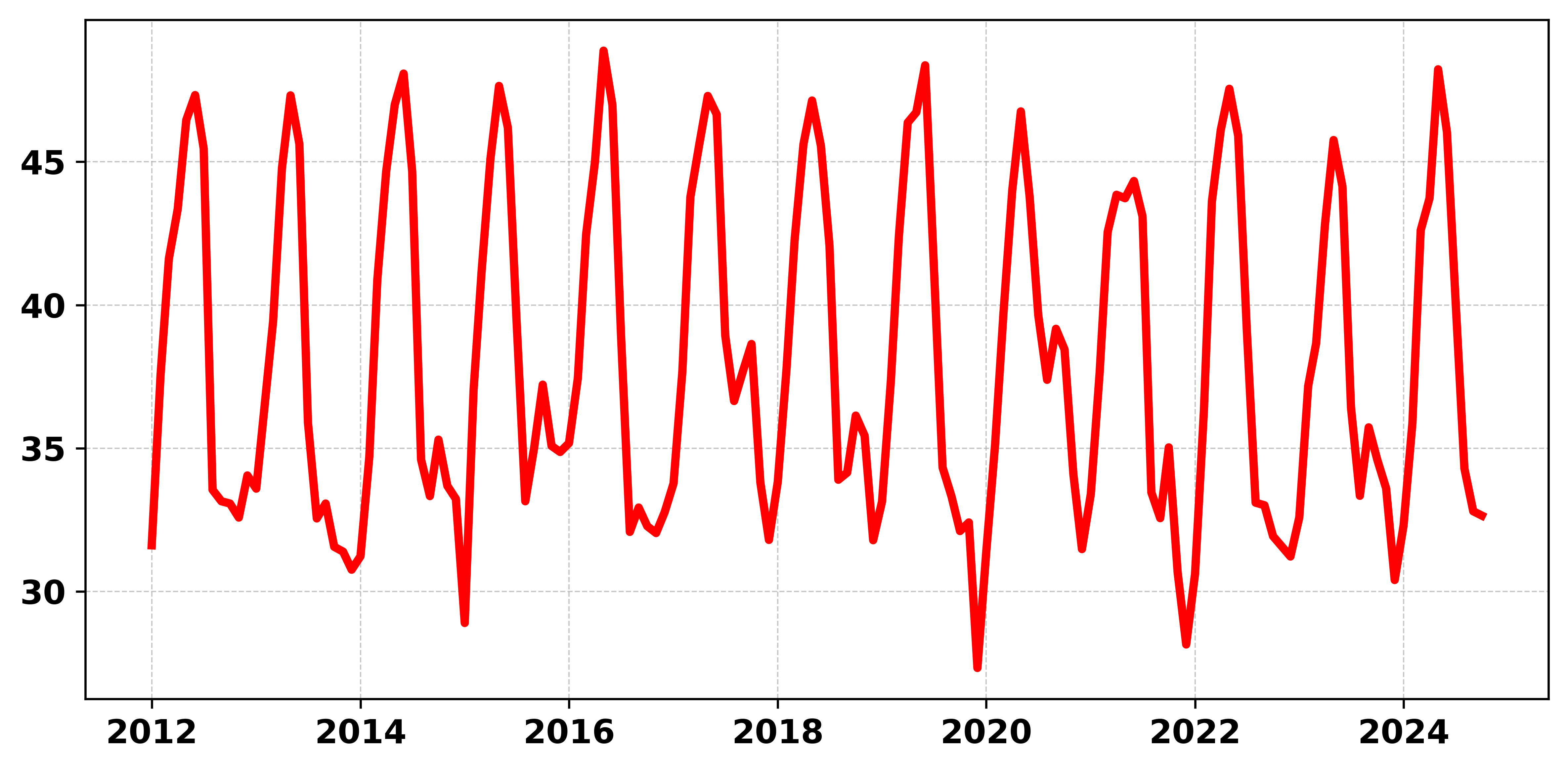}};
  \node[rotate=90] at (-4.5,0) {Maximum Temperature (\textdegree{}C)};
  \node (img4) at (8,0) {
    \includegraphics[width=0.4\linewidth]{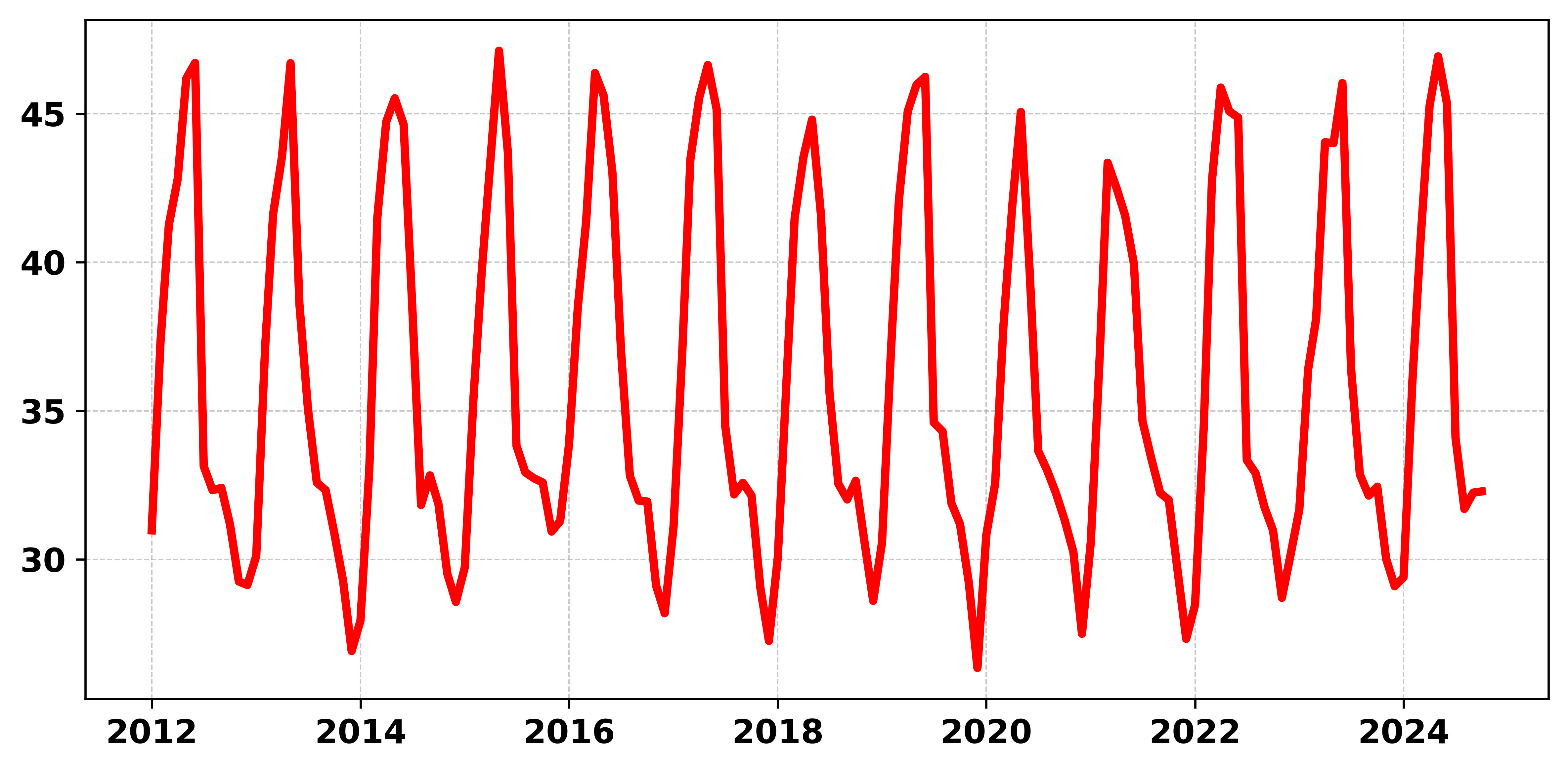}};
\end{tikzpicture}

\captionsetup{justification=justified} 
\caption{\textbf{Meteorological Variables}. The blue and red lines represent the monthly precipitation and maximum temperature from 2012 to 2024 respectively for Madhya Pradesh (left column) and Odisha (right column).}
\label{fig:Prep_Temp_visualizations}
\end{figure}

Figure~\ref{fig:Prep_Temp_visualizations} displays monthly precipitation and maximum temperature for Madhya Pradesh and Odisha. All four plots exhibit pronounced annual seasonal patterns. The precipitation chart (in blue) show significant spikes corresponding to India's monsoon season, when rainfall is highest. The maximum temperature plots (in red) demonstrate clear yearly cycles, with peak temperatures exceeding 40\textdegree{}C.

Table~\ref{tab:statistical_summary} shows the statistical summary for soybean and brinjal monthly prices.
Interestingly, soybean log returns show a negative skew of -0.9839, indicating a tendency for extreme negative returns, which may signal potential downside risks. In contrast, brinjal log returns exhibit a positive skew of 0.3889, reflecting a slight tendency for extreme positive returns, which could indicate more upside potential. The kurtosis value for soybean log returns is 8.0946, indicating a leptokurtic distribution characterized by a sharper peak and fatter tails, suggesting a higher likelihood of extreme values. Conversely, brinjal log returns have a kurtosis of 0.2779, indicating a flatter distribution with less probability of extreme returns.

\begin{table}[H]
    \centering
    \caption{Statistical Summary for Soybean and Brinjal Prices and their Log Returns}
    \begin{tabular}{@{}lcccccc@{}}
        \toprule
        \multirow{2}{*}{Parameter} & \multicolumn{2}{c}{Soybean} & \multicolumn{2}{c}{Brinjal} \\ 
        \cmidrule(lr){2-3} \cmidrule(lr){4-5}
                                   & Price & Log Returns & Price & Log Returns \\ \midrule
        Observations               & 154    & 153        & 154   & 153         \\
        Mean                       & 3867.74 & 0.0040    &  2285.16 & 0.0077    \\
        Standard Deviation         & 1150.03 & 0.0905    & 744.22 &  0.2736    \\
        Minimum                    & 2269.71 & -0.5164   & 1045.83 & -0.7154   \\
        Maximum                    & 8511.76 & 0.3115    & 4250.00 & 0.7862    \\
        Skewness                   & -       & -0.9839   & -      & 0.3889    \\
        Kurtosis                   & -       & 8.0946    & -      & 0.2779    \\ \bottomrule
    \end{tabular}
    \label{tab:statistical_summary}
\end{table}

\section{Methodology and Results}
\subsection{EGARCH Model for Price Volatility Estimation}

We begin our analysis by fitting the Exponential Generalized Autoregressive Conditional Heteroskedasticity (EGARCH) model to the log-returns of soybean and brinjal price data across Madhya Pradesh and Odisha respectively.

\subsubsection{Methodology}
The EGARCH model, an extension of the GARCH family, is particularly suited to capture asymmetric volatility in time-series data. It models the conditional variance, \(\sigma_t^2\), as an asymmetric function of lagged disturbances defined by:
\[
\ln \sigma_t^2 = \omega + \sum_{i=1}^{p} \alpha_i \left( \left| \frac{\epsilon_{t-i}}{\sigma_{t-i}} \right| - \sqrt{\frac{2}{\pi}} \right) + \sum_{j=1}^{o} \gamma_j \frac{\epsilon_{t-j}}{\sigma_{t-j}} + \sum_{k=1}^{q} \beta_k \ln \sigma_{t-k}^2.
\]

Here,
\begin{itemize}
    \item \(\omega\): The constant term represents the baseline level of the log conditional variance.
    \item \(\alpha_i\) (for \(i=1,\dots,p\)): These parameters capture the impact of past shocks on current volatility. The term 
    \[
    \left| \frac{\epsilon_{t-i}}{\sigma_{t-i}} \right| - \sqrt{\frac{2}{\pi}}
    \]
    represents the centered absolute standardized residual, where \(\sqrt{\frac{2}{\pi}}\) is subtracted so that its expected value is zero under a normal distribution.
    \item \(\gamma_j\) (for \(j=1,\dots,o\)): These coefficients measure the asymmetric or leverage effect. They allow the model to differentiate between the impacts of positive and negative shocks on volatility.
    \item \(\beta_k\) (for \(k=1,\dots,q\)): These parameters model the persistence in volatility by incorporating the effect of past conditional variances, expressed in logarithmic form.
    \item \(\epsilon_{t}\): The residuals (or innovations) from the mean equation, representing the unexpected shocks at time \(t\).
    \item \(\sigma_t\): The conditional standard deviation (volatility) at time \(t\), with \(\sigma_t^2\) being the conditional variance.
    \item \(p\), \(o\), \(q\): The orders of the model corresponding to the number of lagged terms for the absolute standardized shocks, the asymmetric shocks, and the lagged log variances, respectively.
\end{itemize}

\subsubsection{Results}

\begin{figure}[H]
\centering
\begin{tikzpicture}
  \node (img5) at (0,0) {
    \includegraphics[width=0.75\linewidth]{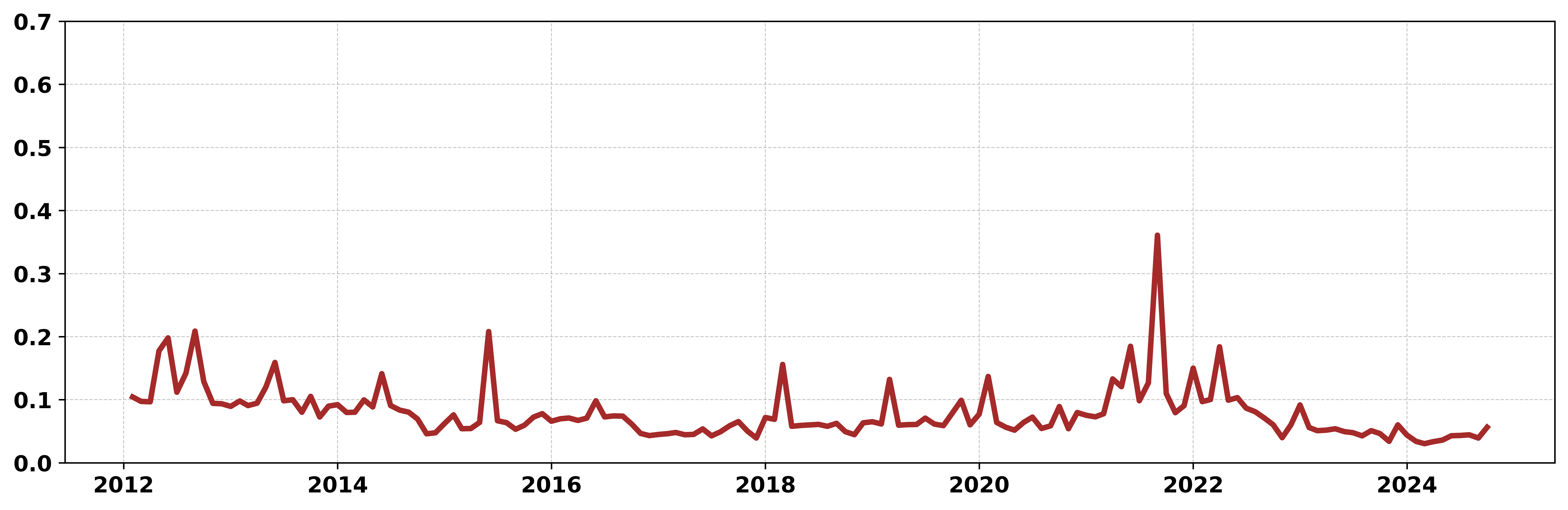}};
  \node[rotate=90] at (-7,0) {Conditional Volatility};
  \end{tikzpicture}
  \begin{tikzpicture}
    \node (img6) at (0,0) {
    \includegraphics[width=0.75\linewidth]{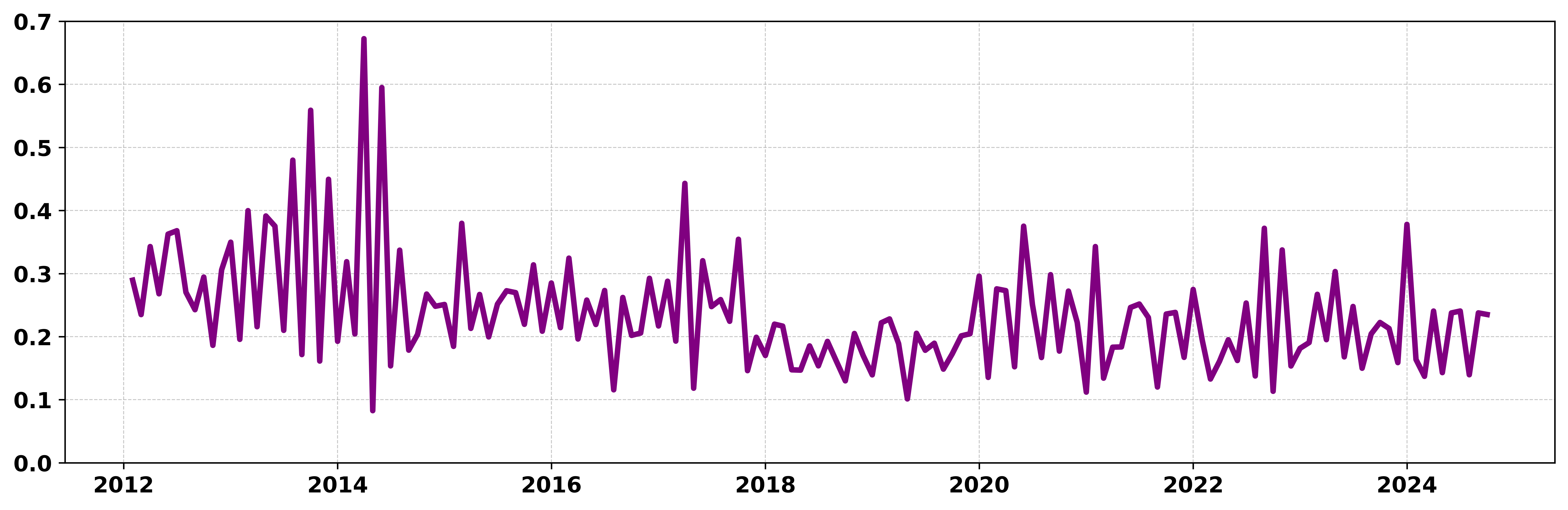}};
  \node[rotate=90] at (-7,0) {Conditional Volatility};
  \node[rotate=0] at (0,-2.5) {Year};
\end{tikzpicture}

\captionsetup{justification=justified} 
\caption{\textbf{Conditional Volatility}. The brown and purple lines represent the conditional volatility of monthly log returns for soybean and brinjal prices respectively.}
\label{fig:cV_visualizations}
\end{figure}

Figure~\ref{fig:cV_visualizations} illustrates the conditional volatility of monthly log returns using the EGARCH model for soybean and brinjal prices. A notable spike in soybean price volatility is observed in late 2021, coinciding with the lingering effects of COVID-19 pandemic. As an export-oriented oilseed, soybean was heavily affected by the aftereffects of the lockdown, impacting its supply chain resulting in extreme volatility in their prices. Being a kharif crop, it is typically sown during the onset of monsoon in the months of June-July and harvested during October. Hence, volatility spikes are observed primarily during specific periods, likely due to harvest and supply fluctuations. Outside these periods, price volatility movements remain relatively stable due to consistent supply. In contrast, brinjal does not have a specific harvesting season therefore exhibits persistent volatility throughout the year. A sharp increase and decline in brinjal prices can be observed during 2014 (April-May), when delayed rainfall led to sharp increase in brinjal prices, followed by a rapid decrease as rainfall improved and supply stabilized\cite{businessstandard2014}.

Table~\ref{tab:EGARCH_Parameter_Estimates} presents the parameter estimates obtained from the EGARCH model applied to crop price log returns.

\begin{table}[H]
    \centering
    \caption{Parameter Estimates of EGARCH model for Soybean and Brinjal}
    \begin{tabular}{@{}lcc|cc@{}}
        \toprule
        \multirow{2}{*}{Parameter} & \multicolumn{2}{c|}{Soybean} & \multicolumn{2}{c}{Brinjal} \\ 
        \cmidrule(lr){2-3} \cmidrule(lr){4-5}
                                    & Estimate & Std. Err & Estimate & Std. Err \\ 
        \midrule
        $\mu$ (Mean)                & 0.00877  & 0.00618  & 0.03656  & 8.559e-07 \\
        $\omega$ (Constant)         & -0.41060 & 0.11445 & -0.22599 & 2.202e-12 \\
        $\alpha_1$ (ARCH Coefficient) & 0.10250  & 0.16198 & 0.64441 & 1.002e-10 \\
        $\alpha_2$ (ARCH Coefficient) & 0.15469 & 0.21614 & -1.06278 & 3.357e-08 \\
        $\alpha_3$ (ARCH Coefficient) & -0.37506 & 0.16464 & 0.15620 & 1.830e-09 \\
        $\gamma_1$ (Leverage Effect)  & 0.01845 & 0.13879 & -0.06083 & 8.479e-07 \\
        $\gamma_2$ (Leverage Effect)  & 0.52901 & 0.17544 & - & - \\
        $\gamma_3$ (Leverage Effect)  & -0.24236  & 0.17653 & - & - \\
        $\beta_1$ (GARCH Coefficient) & 0.92453  & 0.02249 & 0.29258 & 1.026e-10 \\
        $\beta_2$ (GARCH Coefficient) & -       & - & 0.62815 & 6.449e-10 \\
        \midrule
    \end{tabular}
    \label{tab:EGARCH_Parameter_Estimates}
\end{table}

\subsection{Cross-Correlation and Causality}
\subsubsection{Methodology}
To understand the dynamic relationship between conditional volatility and meteorological variables (maximum temperature and precipitation) we calculate the cross-correlation between them at different lags. Cross-correlation analysis is a statistical tool that measures the correlation between two time series as a function of the time-lag applied to one of them. This technique allows us to quantify the extent to which past climate conditions influence current volatility, which is critical for forecasting future price movements based on meteorological trends.

Given two time series: the first representing the conditional volatility obtained from EGARCH, denoted as \( X_t \), and the second representing the meteorological variables (maximum temperature or precipitation), denoted as \( Y_t \), the cross-correlation function (CCF) is computed as:
\[
\text{CCF}(k) = \frac{\sum_{t=k+1}^{T} (X_t - \bar{X})(Y_{t-k} - \bar{Y})}{\sqrt{\sum_{t=k+1}^{T}(X_t - \bar{X})^2 \sum_{t=k+1}^{T}(Y_{t-k} - \bar{Y})^2}},
\]
where \( k \) is the lag applied to the second time series \( Y_t \), \( \bar{X} \) and \( \bar{Y} \) are the mean values of the time series \( X_t \) and \( Y_t \), respectively, and \( T \) is the length of the time series. The cross-correlation function measures the correlation of volatility on meteorological variables at different lag values. CCF indicates the relationship between two variables, however, it does not comment anything of their cause-effect relation or whether one variable holds the information to forecast another. To understand this relationship, the Granger-causality test is required.

Granger-causality test is a statistical test to identify the cause and effect of one time series on another~\cite{Granger1969,sharma2017complex}. We estimate the Granger-causality between volatility and past meteorological variables to analyze the cause of the meteorological variables on the volatility.
In order to estimate the cause of time series $Y$ on time series $X$, a regression is performed. $X$ is modeled from it's own lagged values and the lagged values of $Y$. Further, we evaluate the significance of coefficient linked with $Y$ to check if they are useful for forecasting $X$~\cite{}.

\begin{equation}
    X_t = \alpha_0 + \alpha_1 X_{t-1} + ... + \alpha_k X_{t-k} + \beta_1 Y_{t-1} + ... + \beta_k Y_{t-k} + \epsilon_t,
\end{equation}
where $k$ is the no. of lags, $\alpha$ and $\beta$ are the coefficients of the lagged values of $X$ and $Y$, respectively, and $\epsilon$ is the prediction error. Granger-causality employs hypothesis testing to evaluate the significance of the coefficients by computing the $p$ value~\cite{Granger1969,sharma2017complex}. The null hypothesis states that $X$ does not cause $Y$ indicating that the lagged values does not hold any information to improve the forecasting.We reject the null hypothesis if the $p$ value is less than 0.05.

\subsubsection{Results}

\begin{figure}[H]
\centering
\begin{tikzpicture}
    \node (image1) at (0,0) {\includegraphics[width=.75\linewidth]{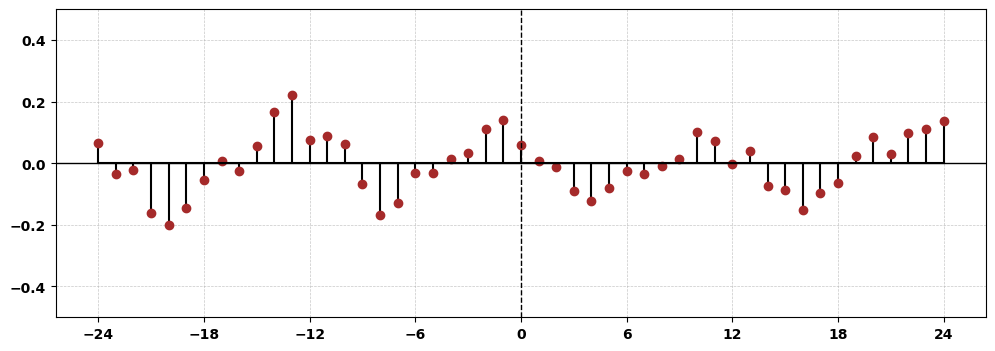}};
    \node[rotate=90] at (-7,0) {Cross-Correlation};
    \end{tikzpicture}
    \begin{tikzpicture}
      \node (image2) at (0,0) {\includegraphics[width=.75\linewidth]{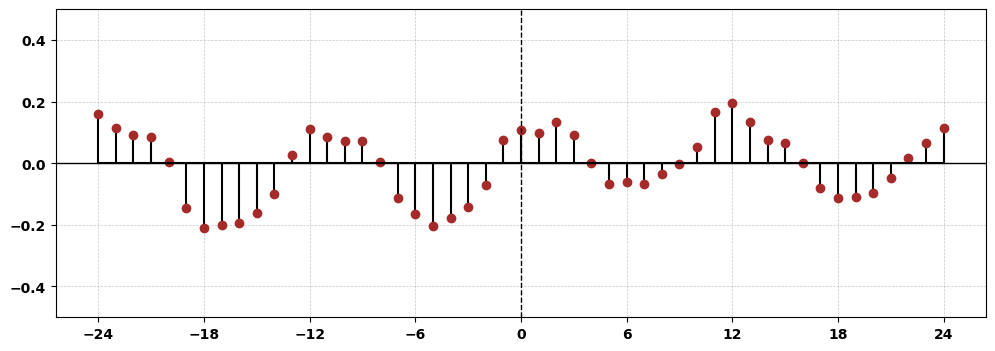}};
    \node[rotate=90] at (-7,0) {Cross-Correlation};
    \node[rotate=0] at (0,-2.5) {Monthly Lag};
\end{tikzpicture}
\captionsetup{justification=justified} 
\caption{\textbf{Cross-Correlation Analysis of soybean price volatility in Madhya Pradesh}. The plot represents the cross-correlation analysis with 24 monthly lags for soybean price volatility; the first plot represents the cross-correlation between volatility and precipitation, while the second plot represents the cross-correlation between volatility and maximum temperature.}
\label{fig:Cross-Correlation_MP}
\end{figure}

\begin{figure}[H]
    \centering
    \begin{tikzpicture}
        \node (image1) at (0,0) {\includegraphics[width=.75\linewidth]{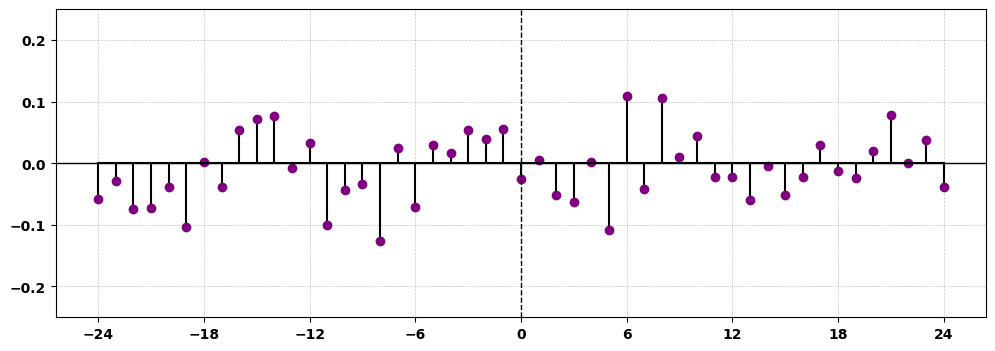}};
        \node[rotate=90] at (-7,0) {Cross-Correlation};
        \end{tikzpicture}
        \begin{tikzpicture}
            
        \node (image2) at (0,0) {\includegraphics[width=.75\linewidth]{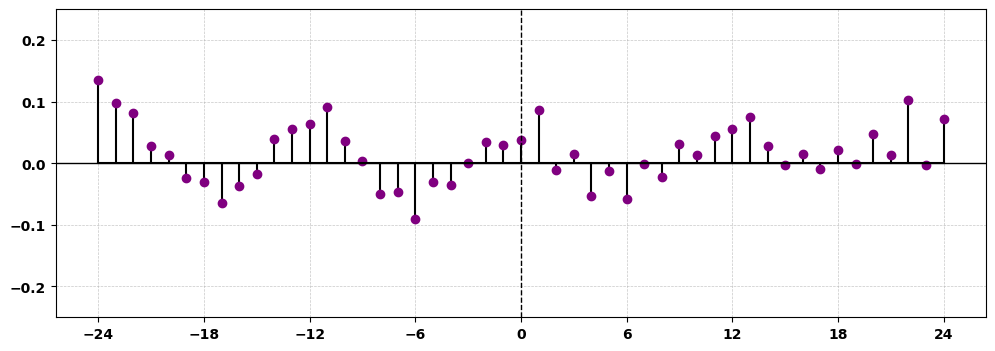}};
        \node[rotate=90] at (-7,0) {Cross-Correlation};
        \node[rotate=0] at (0,-2.5) {Monthly Lag};
    \end{tikzpicture}
    \captionsetup{justification=justified} 
    \caption{\textbf{Cross-Correlation Analysis of brinjal price volatility in Odisha}. The plot represents the cross-correlation analysis with 24 monthly lags for brinjal price volatility; the first plot represents the cross-correlation between volatility and precipitation, while the second plot represents the cross-correlation between volatility and maximum temperature.}
    \label{fig:cross-correlation_OD}
\end{figure}

Figure~\ref{fig:Cross-Correlation_MP} and ~\ref{fig:cross-correlation_OD} showcase the cross-correlation of crop price volatility and meteorological variables for soybean in Madhya Pradesh and brinjal in Odisha respectively. In both cases, we observe a cyclical relationship highlighting the presence of seasonality. The periodic nature of the correlation suggests that weather patterns influence price volatility in a recurring manner. However, the strength of the correlation remains relatively low, with values not exceeding 0.25 or dropping below -0.25 for soybean and remaining within the range of 0.1 to -0.1 for brinjal. This indicates that while meteorological variables exhibit a seasonal impact on volatility, their direct influence is not strongly pronounced, particularly in the case of brinjal.

After estimating the cross-correlation function, it is important to analyze the cause-effect relationship between volatility and meteorological factors. In order to check for a causal relation between volatility and meteorological variables, we conducted a Granger-causality test for both price volatility time series. The stationarity of volatility, maximum temperature, and precipitation was examined using the Kwiatkowski-Phillips-Schmidt-Shin (KPSS) test for both cases. The results confirmed that the time series for soybean price volatility is stationary, whereas brinjal price volatility is not. To address this, the first difference of brinjal price volatility was taken to achieve stationarity, after which we proceeded with the Granger causality test to examine whether maximum temperature and precipitation have an effect on price volatility.

The analysis did not reveal any significant casual relationship between meteorological variables and soybean price volatility in Madhya Pradesh. This can be attributed to the fact that soybean is a globally traded commodity and its prices are largely influenced by international demand and supply conditions. As a result local climate change factors in Madhya Pradesh have minimal impact, which is reflected in the absence of a Granger causal relationship between soybean price volatility and meteorological factors. In contrast, brinjal, a vegetable primarily produced in Odisha, shows a different pattern. In this case, price volatility is related to local demand and supply, which are more sensitive to regional weather conditions. We observe a Granger casual relationship between precipitation and differenced price volatility at lag 6 and 7. This indicates that rainfall patterns significantly influence the market stability of brinjal in Odisha. 

\subsection{Predictive Modelling}
\subsubsection{Methodology}
The next stage of our analysis focuses on developing predictive models that forecast price volatility by incorporating meteorological inputs. To achieve this, we employ two distinct machine learning  approaches\cite{shumway2017time,murphy2012machine}: the \textbf{SARIMAX (Seasonal Autoregressive Integrated Moving Average with Exogenous variables)} model and the \textbf{LSTM (Long Short-Term Memory)} neural network. Each method offers unique advantages in capturing the temporal dynamics and complex relationships between climatic variables and agricultural price volatility.

\paragraph{SARIMAX Model:}

The SARIMAX model expands upon the traditional ARIMA framework by incorporating both seasonal components and external variables, making it particularly well-suited for agricultural price analysis. The general form of the SARIMAX model can be written as:
\[
\Phi_P(L^s) \phi_p(L) (1 - L)^d (1 - L^s)^D \sigma_t = \Theta_Q(L^s) \theta_q(L) \varepsilon_t + \mathbf{x}_t \boldsymbol{\beta},
\]
where:
\begin{itemize}
    \item  \(\Phi_P(L^s)\): Seasonal autoregressive operator of order \(P\), defined as \(\Phi_P(L^s) = 1 - \Phi_1 L^s - \Phi_2 L^{2s} - \dots - \Phi_P L^{Ps}\).
    \item \(\phi_p(L)\): Non-seasonal autoregressive operator of order \(p\), defined as \(\phi_p(L) = 1 - \phi_1 L - \phi_2 L^2 - \dots - \phi_p L^p\).
    \item \(d\): Order of non-seasonal differencing.
    \item \(D\): Order of seasonal differencing.
    \item \(s\): Seasonal period (e.g., \(s = 12\) for monthly data with yearly seasonality).
    \item \(\Theta_Q(L^s)\): Seasonal moving average operator of order \(Q\).
    \item \(\theta_q(L)\): Non-seasonal moving average operator of order \(q\).
    \item \(\varepsilon_t\): Error term.
    \item \(\mathbf{x}_t\): Vector of exogenous regressors (e.g., maximum temperature \(tasmax\) and precipitation \(pr\)), where \texttt{tasmax} is the maximum atmospheric temperature, and \texttt{pr} is precipitation.
    \item \(\boldsymbol{\beta}\): Coefficient vector for exogenous regressors.
    
\end{itemize}

The model captures the linear relationships between the dependent variable (price volatility) and both its own past values and external meteorological variables while also accounting for seasonal patterns. By incorporating these exogenous variables, the model can predict future volatility based on past patterns in the data and concurrent climate conditions. The model fitting involves selecting appropriate values for \( p \), \( q \), \( d \), \( P \), \( Q \) and \( D \), where \( d \) and \( D \) represent the differencing and seasonal differencing order respectively, to make the time series stationary when needed. To determine the optimal model parameters, we optimized the Akaike Information Criterion (AIC)\cite{murphy2012machine}, which helped balance model complexity with predictive accuracy.

\paragraph{LSTM Model:}

To capture non-linear and complex dynamics between climatic variables and price volatility, an LSTM model is used. The LSTM model is a type of recurrent neural network (RNN)\cite{shumway2017time,murphy2012machine} that is designed to handle long-term dependencies in sequential data, making it well-suited for time series forecasting in agricultural markets, where the effects of weather patterns on prices may exhibit delayed interactions.

The model consists of a series of memory cells, each containing three main components: the input gate, the forget gate, and the output gate. These gates control the flow of information into and out of the memory cells, allowing the network to "remember" important information for long periods of time and "forget" irrelevant information. The mathematical formulation of an LSTM unit is as follows:

The Forget Gate in an LSTM model regulates which past information is retained or discarded in the memory cell, computed as 
\[
f_t = \sigma(W_f \cdot [h_{t-1}, x_t] + b_f),
\]
where \( f_t \) is the forget gate vector, \( \sigma \) is the sigmoid activation function, \( W_f \) are the weight matrices, \( h_{t-1} \) is the previous hidden state, \( x_t \) is the current input, and \( b_f \) is the bias. The Input Gate determines what new information should be updated into the memory, calculated as
\[
i_t = \sigma(W_i \cdot [h_{t-1}, x_t] + b_i) \quad \text{and} \quad \tilde{C}_t = \tanh(W_C \cdot [h_{t-1}, x_t] + b_C)
\]
where \( i_t \) is the input gate vector and \( \tilde{C}_t \) is the candidate cell state. The cell state is updated as 
\[
C_t = f_t \cdot C_{t-1} + i_t \cdot \tilde{C}_t,
\]
combining the previous state with the new information. Finally, the Output Gate produces the final output based on the updated cell state, calculated as 
\[
o_t = \sigma(W_o \cdot [h_{t-1}, x_t] + b_o) \quad \text{and} \quad h_t = o_t \cdot \tanh(C_t)
\]
where \( o_t \) is the output gate vector, and \( h_t \) is the hidden state (the final output of the LSTM unit at time \( t \)).The LSTM model is trained on sequences of past price volatility and meteorological data (maximum temperature and precipitation). Its architecture enables it to learn both short-term and long-term dependencies, making it highly effective in capturing the delayed and intricate effects of climate variability on agricultural price volatility.

\subsubsection{Results}

\begin{figure}[H]
    \centering
    \begin{tikzpicture}
        \node (image1) at (0,0) {\includegraphics[width=.75\linewidth]{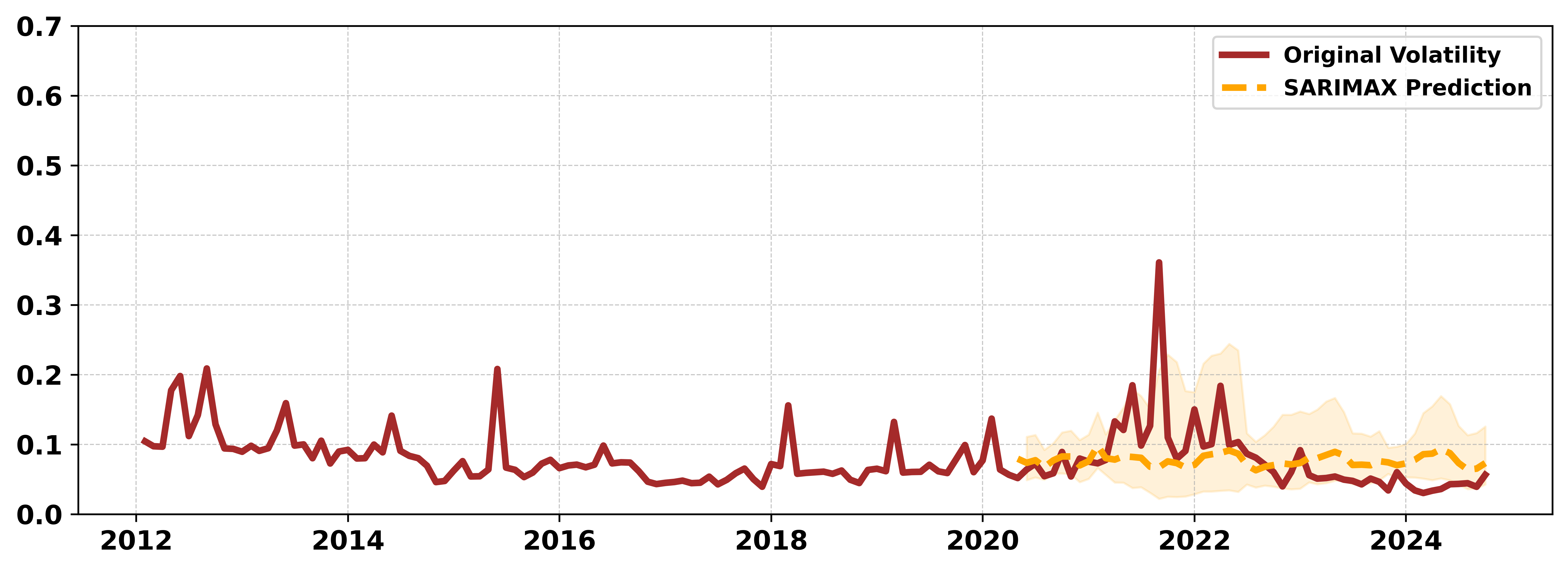}};
        \node[rotate=90] at (-7,0) {Conditional Volatility};
        \end{tikzpicture}
        \begin{tikzpicture}
         \node (image2) at (0,0) {\includegraphics[width=.75\linewidth]{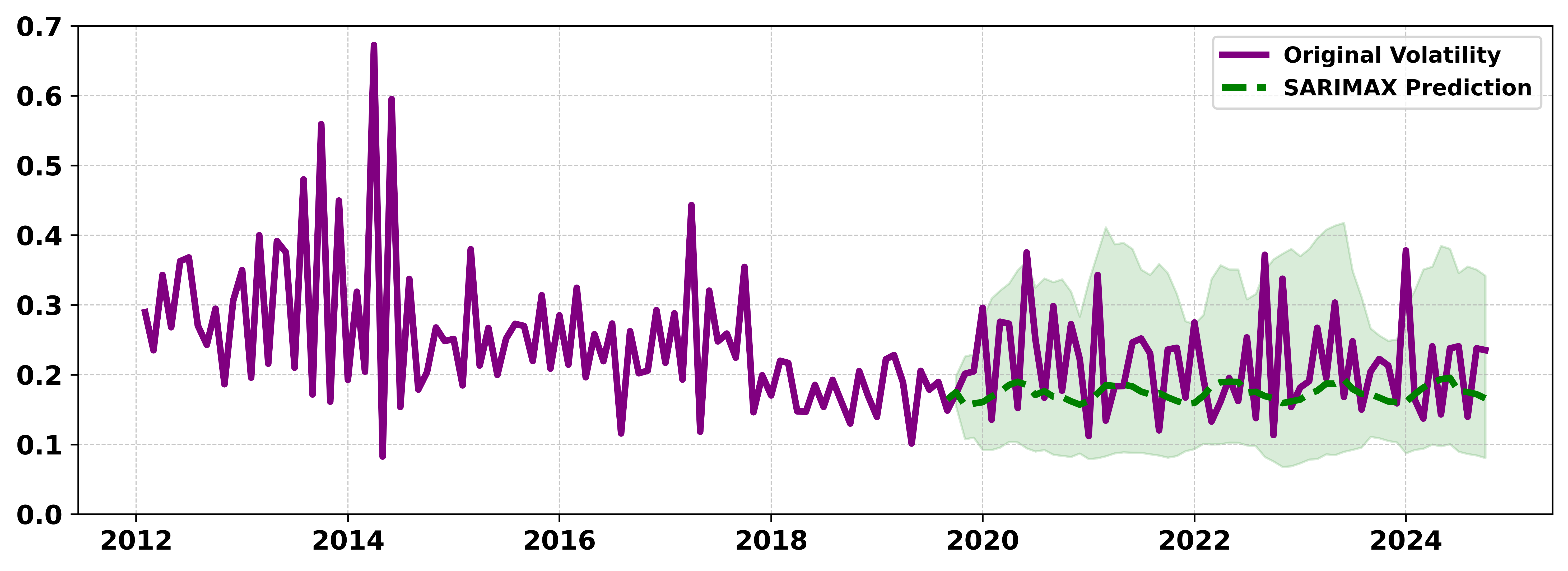}};
        \node[rotate=90] at (-7,0) {Conditional Volatility};
        \node at (0,-2.5) {Year};
    \end{tikzpicture}
    \captionsetup{justification=justified}
    \caption{\textbf{Comparison of SARIMAX Forecasts}. 
The brown and purple lines represent the conditional volatility of log-returns for soybean prices in Madhya Pradesh and brinjal prices in Odisha respectively. In first plot, the orange dashed line represents the SARIMAX prediction for soybean price volatility during the testing period from May 2020 to October 2024.  
Similarly, in second plot, the green dashed line represents the SARIMAX prediction for brinjal price volatility over the testing period of September 2019 to October 2024.}
\label{fig:sarimax_comparison}
\end{figure}

Figure~\ref{fig:sarimax_comparison} compares the conditional volatility obtained from the EGARCH model with the SARIMAX-predicted volatility for soybean and brinjal, respectively. The results indicate that the forecasting error is higher for the price volatility of soybean. This may be because the volatility during the testing period (Sep 2019 - Oct 2024) was more driven by the external shocks such as COVID-19 than meteorological factors. As previously discussed, soybean, being an export-oriented crop, is more susceptible to global disruptions. In contrast, brinjal, which is primarily domestically consumed, is less affected by such external influences.

\begin{figure}[H]
    \centering
       \begin{tikzpicture}
        \node (image1) at (0,0) {\includegraphics[width=.75\linewidth]{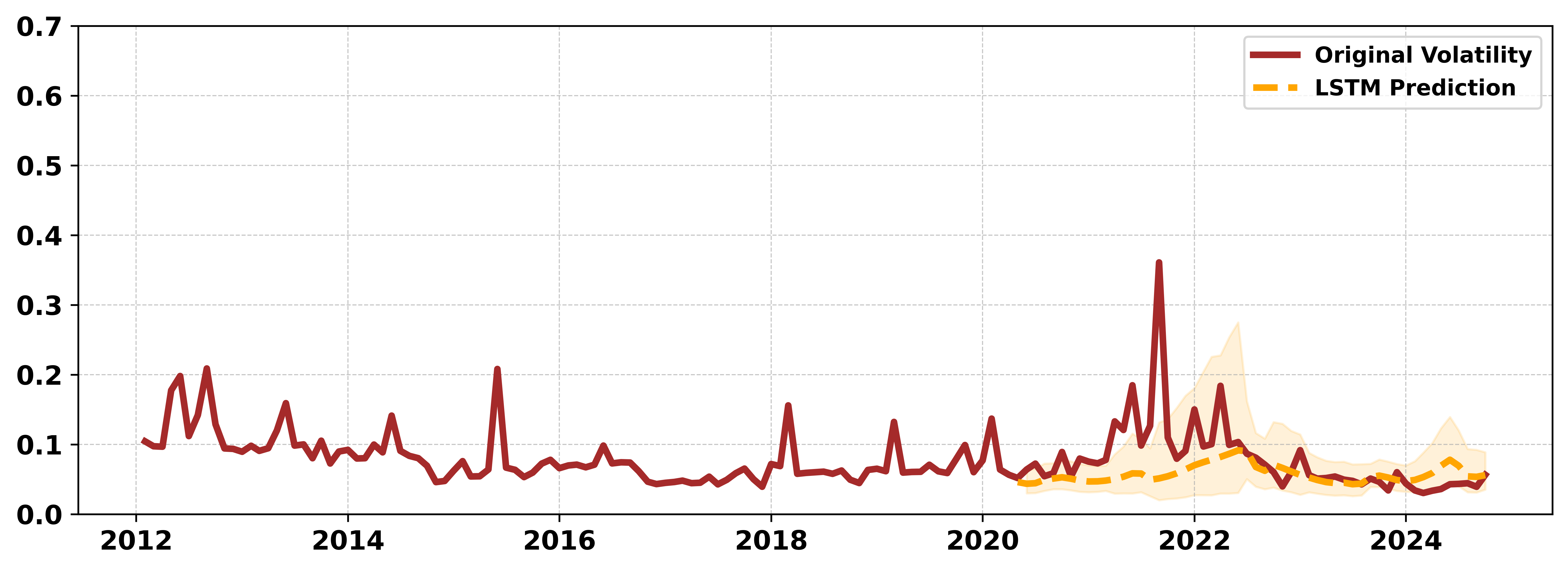}};
        \node[rotate=90] at (-7,0) {Conditional Volatility};
        \end{tikzpicture}
        \begin{tikzpicture}
         \node (image2) at (0,0) {\includegraphics[width=.75\linewidth]{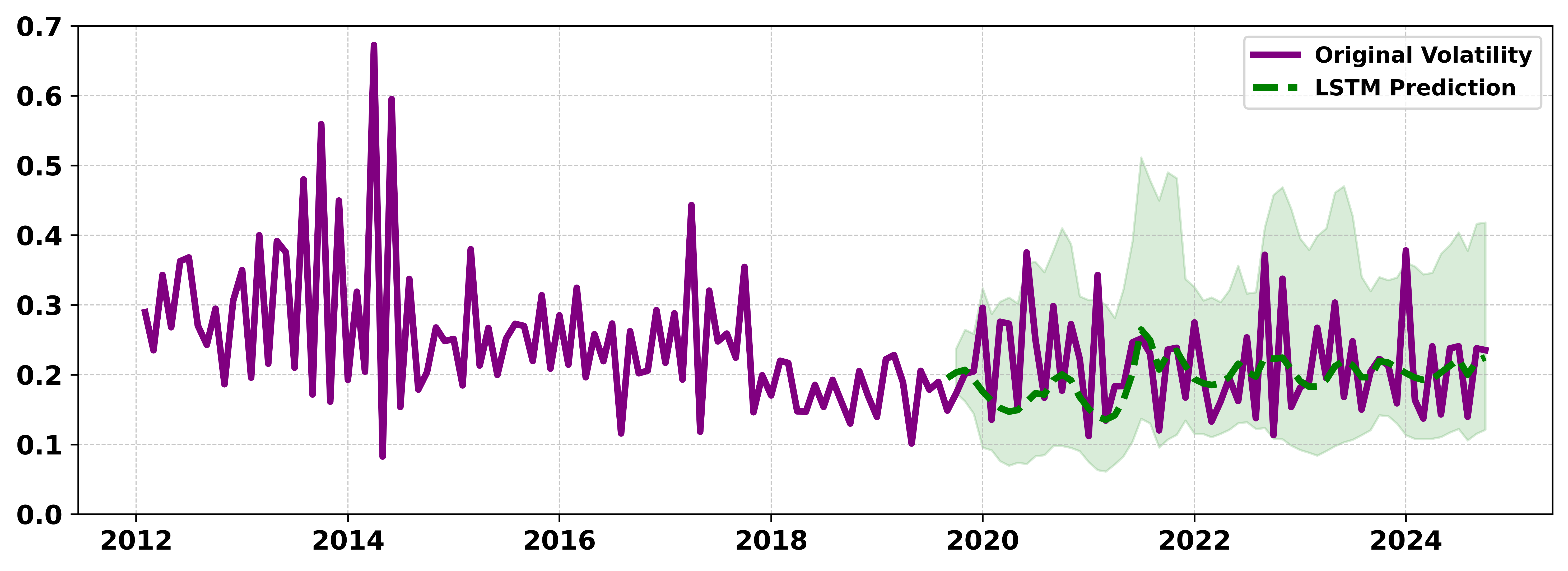}};
        \node[rotate=90] at (-7,0) {Conditional Volatility};
        \node at (0,-2.5) {Year};
    \end{tikzpicture}
    
    \captionsetup{justification=justified}
    \caption{\textbf{Comparison of LSTM Forecasts}.The brown and purple lines represent the conditional volatility of log returns for soybean prices in Madhya Pradesh and brinjal prices in Odisha respectively. In first plot, the orange dashed line represents the LSTM model’s prediction for soybean price volatility during the testing period from May 2020 to October 2024.  
    Similarly, in second plot, the green dashed line represents the LSTM prediction for brinjal price volatility over the testing period of September 2019 to October 2024.}
    \label{fig:lstm_comparison}
\end{figure}

Similarly, Fig.~\ref{fig:lstm_comparison} compares the conditional volatility obtained from the EGARCH model with the LSTM-predicted volatility for soybean and brinjal, respectively. As discussed earlier, the higher forecast error for soybean price volatility is likely due to external disruptions such as COVID-19 playing a more dominant role, whereas brinjal, being less affected by global shocks, exhibits a lower prediction error.

\begin{table}[h]
    \centering
    \renewcommand{\arraystretch}{1.2}
    \begin{tabular}{ll|l|c}
        \hline
        \textbf{State} & \textbf{Crop} & \textbf{Model} & \textbf{MAPE} \\ 
        \hline
        \multirow{2}{*}{Madhya Pradesh} & \multirow{2}{*}{Soybean} 
            & SARIMAX   & 0.48 \\ \cline{3-4}
            &  & LSTM     & 0.33 \\ 
        \hline
        \multirow{2}{*}{Odisha} & \multirow{2}{*}{Brinjal} 
            & SARIMAX  & 0.26 \\ \cline{3-4}
            &  & LSTM     & 0.24 \\
        \hline
    \end{tabular}
    \caption{MAPE Comparison for SARIMAX and LSTM Models}
    \label{tab:model_comparison}
\end{table}

\subsection{Spatio-Temporal Volatility Estimation}
\subsubsection{Methodology}
For the final step of our analysis, we study the spatial distribution of conditional volatility in soybean prices across Madhya Pradesh and brinjal prices across Odisha. As such, we construct a series of monthly volatility surfaces using a combination of spatial statistical methods. 

For each month, we extract district-level crop price conditional volatility estimates and compute a spatial neighborhood structure using the k-Nearest Neighbors (KNN) algorithm~\cite{murphy2012machine}, selecting the two nearest neighbors for each observation. This structure is converted to a symmetric binary adjacency matrix, representing first-order spatial contiguity--direct connections between geometrically proximate observations. This adjacency matrix serves as the basis for modelling spatial dependency using the Leroux Conditional Autoregressive (CAR) model~\cite{lee2013carbayes}, a Bayesian hierarchical framework that produces smoothed volatility estimates that balance local variation with geographic dependence. These smoothed values were then interpolated across a 0.01-degree resolution grid using Inverse Distance Weighting (IDW)~\cite{webster2007geostatistics}. This deterministic method assigns greater weight to closer observations, enabling the construction of continuous volatility surfaces that capture local heterogeneity while preserving the spatial dependence structure inferred from the CAR model.

\subsubsection{Results}
\begin{figure}[H]
    \centering
     \begin{tikzpicture}
        \node (image1) at (0,0) {\includegraphics[width=1\linewidth]{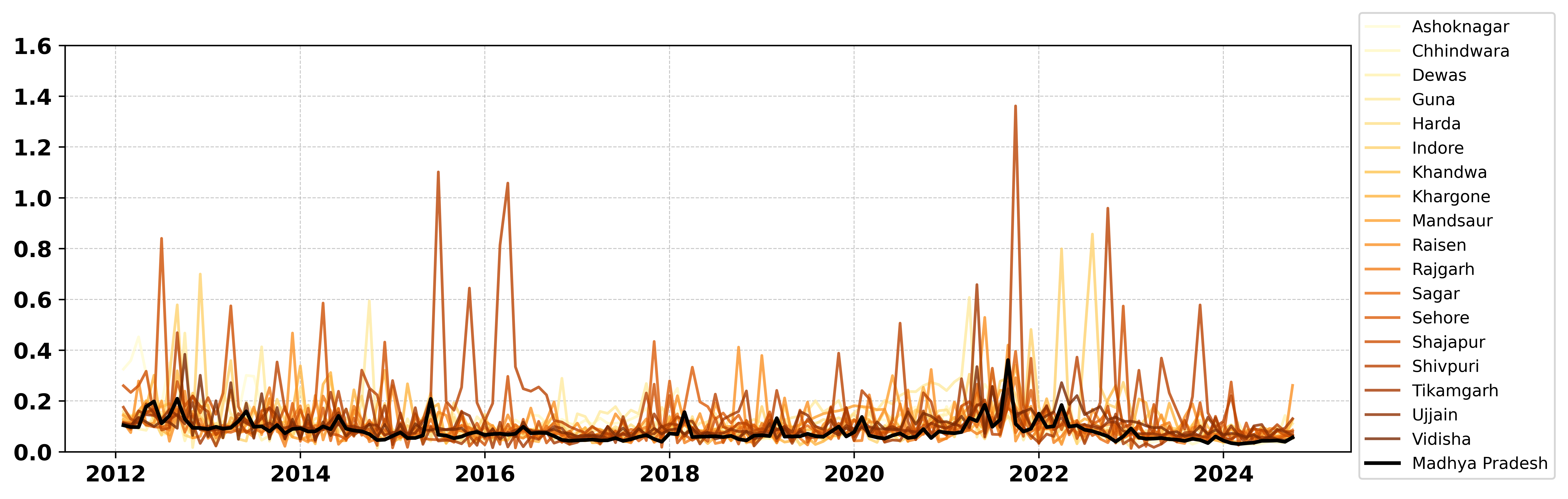}};
        \node[rotate=90] at (-9.3,0) {Conditional Volatility};
     \end{tikzpicture}
     \begin{tikzpicture}
        \node (image2) at (0,0) {\includegraphics[width=1\linewidth]{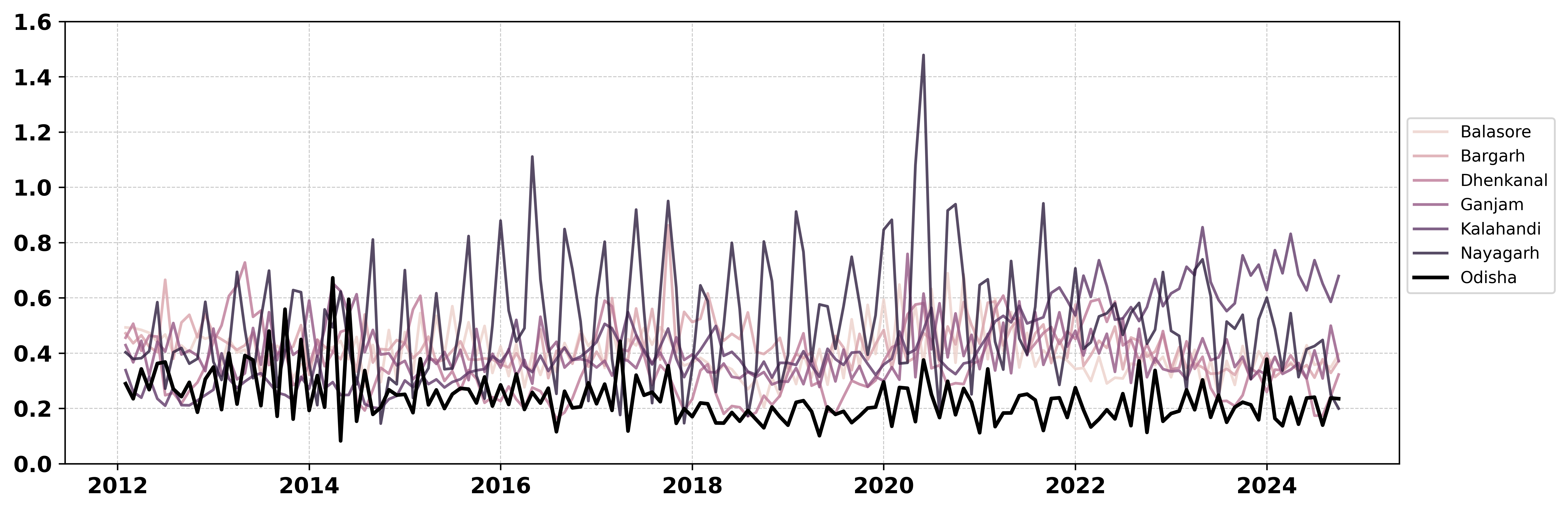}};
        \node[rotate=90] at (-9.3,0) {Conditional Volatility};
        \node at (-0.8,-3) {Year};
     \end{tikzpicture}
    
    \captionsetup{justification=justified}
    \caption{\textbf{District-wise Conditional Volatility}. The brown and purple gradient lines represent the conditional volatility of soybean price in different districts of Madhya Pradesh and brinjal price in different districts of Odisha respectively. The black lines represent the state-wise conditional volatility.}
    \label{fig:districtwise_comparison}
\end{figure}

Figure~\ref{fig:districtwise_comparison} compares the district-wise conditional volatility of the monthly log returns estimated through the EGARCH model for soybean and brinjal prices respectively. It is apparent from the plots that certain districts exhibit substantially higher levels of price volatility compared to others. This suggests the presence of localised factors that can amplify price fluctuations in specific districts.

\begin{figure}[H]
    \centering
     \begin{tikzpicture}
        \node (image1) at (0,0) {\includegraphics[width=1\textwidth, height=0.45\textwidth]{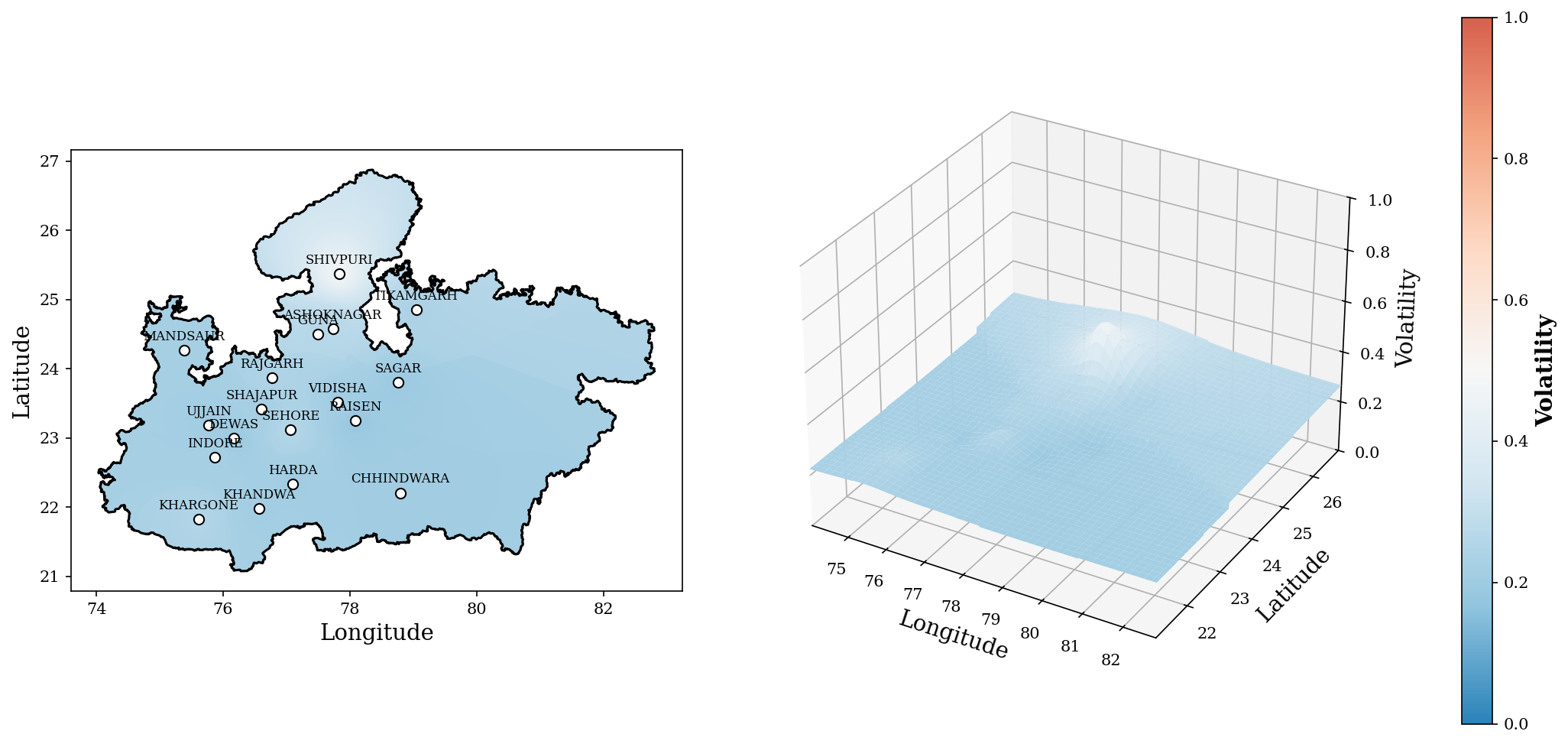}};
     \end{tikzpicture}
     \begin{tikzpicture}
        \node (image2) at (0,0) {\includegraphics[width=1\textwidth, height=0.45\textwidth]{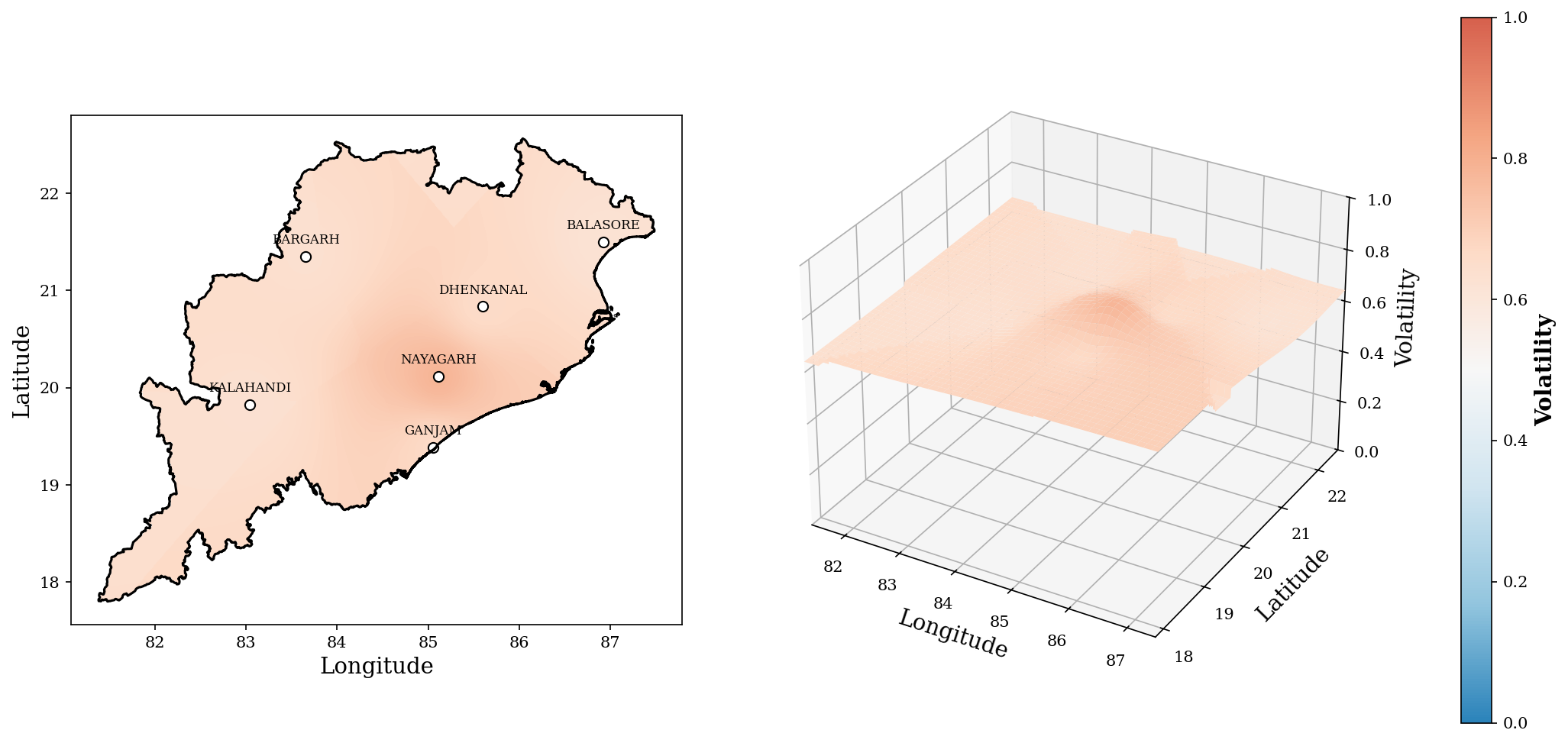}};
     \end{tikzpicture}
     \captionsetup{justification=justified}
     \caption{\textbf{Volatility Surface Estimation using CAR model}. The top row shows the 2D spatial distribution (left) and the corresponding 3D interpolated surface (right) of soybean price volatility in Madhya Pradesh for October 2021. The bottom row presents analogous plots for brinjal price volatility in Odisha for June 2020. The frames were taken from spatial plots available in the Crop Price Analysis app~\cite{abbinavsk_crop-price-analysis}.}
     \label{fig:VolatilitySurface_2D_3D}
\end{figure}

Figure~\ref{fig:VolatilitySurface_2D_3D} presents the volatility surface for soybean price in Madhya Pradesh on October 2021 and brinjal price in Odisha on June 2020. This spatial analysis shows the interpolated distribution of volatility across the state, with higher peaks indicating places of greater uncertainty. The analysis was performed for each time step and the resulting frames were stitched into a GIF, which is displayed in the Crop Price Analysis app~\cite{abbinavsk_crop-price-analysis}.

As mentioned before, soybean price volatility has been increasing across all districts in Madhya Pradesh post-2021 due to shocks caused by the Covid-19 pandemic. In the case of brinjal prices in Odisha, volatility has been noticeably greater post-2020 primarily due to erratic rainfall patterns affecting different districts unevenly~\cite{ndtv2020monsoon}. These climate uncertainties, coupled with localised factors such as inadequate irrigation infrastructure, have further amplified fluctuations in production and market prices, leading to increased volatility at the district level.

\section{Discussions \& Way Forward}

This study examines the impact of meteorological variables on the price volatility of soybean and brinjal, highlighting their differing sensitivities to external disruptions. The EGARCH model revealed that soybean exhibited higher volatility compared to brinjal, primarily due to its export-oriented nature, which made it more susceptible to global shocks such as the COVID-19 pandemic. In contrast, brinjal, being a domestically consumed vegetable, showed relatively stable price volatility patterns. Statistical analysis further supported this, as soybean displayed higher skewness and kurtosis, indicating a greater likelihood of extreme price movements.

To forecast price volatility, SARIMAX and LSTM models were applied using meteorological variables. While LSTM provided better accuracy for soybean, both models performed similarly for brinjal. However, the forecasting accuracy for soybean was lower overall, likely due to heightened volatility in 2020 caused by external disruptions beyond meteorological factors. This suggested that additional variables, such as global trade dynamics and policy interventions, could be incorporated to improve predictive accuracy. To further understand spatial variation in price volatility, the CAR model followed by IDW interpolation was applied monthly, to construct volatility surfaces for each state. These surfaces revealed clear differences in crop price volatility between districts, driven by both common shocks such as the COVID-19 pandemic and localized factors such as inadequate irrigation infrastructure and uneven market access.

Our findings have important implications, particularly for regions which are highly vulnerable to climate change. Improved volatility forecasts could help farmers optimize sowing and harvesting schedules, make informed crop allocation decisions, and mitigate economic risks associated with price fluctuations. An extension of this study can be found in our paper \textit{Predicting and Mitigating Agricultural Price Volatility Using Climate Scenarios and Risk Models}\cite{das2025predicting}, where we examine the complex interactions between crop price volatility, Minimum Support Prices(MSP), and insurance premiums. As agricultural markets become increasingly sensitive to both global shocks and local climate variability, a holistic approach becomes essential to safeguard farmer incomes and ensure food security. Future research could build on this work by integrating additional factors beyond meteorological variables to refine volatility predictions and better understand agricultural price dynamics. Furthermore, the methodology in this study can be replicated for other Indian states and translated into an interactive dashboard to support decision-making by policymakers, researchers, and other stakeholders.   

\section*{Acknowledgements}
The authors thank Himani Gupta for downloading the data of the meteorological factors.

\bibliographystyle{plain}
\bibliography{athesis}

\end{document}